\documentclass[useAMS,usenatbib]{mn2e}
\usepackage{epsfig}

\def\reff@jnl#1{{\rm#1\/}}

\def\aj{\reff@jnl{AJ}}                  
\def\araa{\reff@jnl{ARA\&A}}            
\def\apj{\reff@jnl{ApJ}}                
\def\apjl{\reff@jnl{ApJ}}               
\def\apjs{\reff@jnl{ApJS}}              
\def\ao{\reff@jnl{Appl.Optics}}         
\def\apss{\reff@jnl{Ap\&SS}}            
\def\aap{\reff@jnl{A\&A}}               
\def\aapr{\reff@jnl{A\&A~Rev.}}         
\def\aaps{\reff@jnl{A\&AS}}             
\def\azh{\reff@jnl{AZh}}                
\def\baas{\reff@jnl{BAAS}}              
\def\jrasc{\reff@jnl{JRASC}}            
\def\memras{\reff@jnl{MmRAS}}           
\def\mnras{\reff@jnl{MNRAS}}            
\def\pra{\reff@jnl{Phys.Rev.A}}         
\def\prb{\reff@jnl{Phys.Rev.B}}         
\def\prc{\reff@jnl{Phys.Rev.C}}         
\def\prd{\reff@jnl{Phys.Rev.D}}         
\def\prl{\reff@jnl{Phys.Rev.Lett}}      
\def\pasp{\reff@jnl{PASP}}              
\def\pasj{\reff@jnl{PASJ}}              
\def\qjras{\reff@jnl{QJRAS}}            
\def\skytel{\reff@jnl{S\&T}}            
\def\solphys{\reff@jnl{Solar~Phys.}}    
\def\sovast{\reff@jnl{Soviet~Ast.}}     
\def\ssr{\reff@jnl{Space~Sci.Rev.}}     
\def\zap{\reff@jnl{ZAp}}                
\def\nat{\reff@jnl{Nature}}             
\title
    []
    {Statistical constraints on the IR galaxy number counts and
      cosmic IR background from the Spitzer GOODS survey}

\author
    [Richard S. Savage et al.]
    {Richard S. Savage$^{1}$ \thanks{E-mail: r.s.savage@sussex.ac.uk}, 
     Seb Oliver$^1$ 
     \\
     $^1$ Astronomy Centre, University of Sussex, UK\\}
\begin{document}
\date{Accepted *date*. Received *date*; in original form *date*}
\pagerange{\pageref{firstpage}--\pageref{lastpage}} 
\pubyear{2005}
\maketitle
\label{firstpage}
\begin{abstract}
We perform fluctuation analyses on the data from the Spitzer GOODS
survey (epoch one) in the Hubble Deep Field North (HDF-N).  We fit a
parameterised power-law number count model of the form 
${dN \over dS} =N_o S^{-\delta}$
to data from each of the four Spitzer IRAC bands, using Markov Chain
Monte Carlo (MCMC) sampling to explore the posterior probability
distribution in each case.  We obtain best-fit reduced chi-squared
values of (3.43 0.86 1.14 1.13) in the four IRAC bands. From this
analysis we determine the likely differential faint source counts down
to $10^{-8} Jy$, over two orders of magnitude in flux fainter than has
been previously determined.  

From these constrained number count models, we estimate a lower bound
on the contribution to the Infra-Red (IR) background light arising from faint
galaxies.  We estimate the total integrated background IR light
in the Spitzer GOODS HDF-N field due to faint sources.  By
adding the estimates of integrated light given by
\citet[][]{Fazio-04}, we calculate the total integrated background
light in the four IRAC bands.  We compare our 3.6
micron results with previous background estimates in similar bands and
conclude that, subject to our assumptions about the noise
characteristics, our analyses are able to account for the vast
majority of the 3.6 micron background.  Our analyses are sensitive to
a number of potential systematic effects; we discuss our assumptions
with regards to noise characteristics, flux calibration and
flat-fielding artifacts.  

We compare our results with the galaxy number counts
measured directly by \citet[][]{Fazio-04}.  There is evidence of
a systematic difference between our results and these number counts,
with our fluctuation analysis preferring higher faint number counts.
This could be explained by a relative difference in flux calibration of 50\%.
Such a difference would leave our detection of excess faint number
counts intact, but substantially reduce our estimates of the IR background.
\end{abstract}
\begin{keywords}
Cosmology: diffuse radiation
Galaxies: high-redshift  
Galaxies: statistical  
Infrared: galaxies  
Methods: statistical  
\end{keywords}
\section{Introduction} \label{introduction}
A revolution is occurring in IR astronomy.  With most
wave-bands unobservable from the ground, observations of the IR sky are
heavily reliant on space telescopes.  The launch in August 2003
of the Spitzer space telescope marked the start of a period that will
see as many as five IR satellite telescopes launched by 2010.  With ASTRO-F
(launching in early 2006) \citep[see e.g.][]{ASTROFsuperIRAS-2004},
Herschel (due to launch in 2007) \citep[][]{HerschelOverview-2004}, 
as well as the planned WISE \citep[][]{WISE-2004}
and SPICA \citep[][]{SPICA-2004}
missions, we are entering an epoch of unparalleled access to the IR sky.

IR data are critically important to our understanding of
galaxy formation.  The tight mass:light ratio of
the stellar populations, coupled with relatively minimal dust
obscuration at these wavelengths, make near-IR observations a powerful
tool for probing the mass of stars in a galaxy.

The galaxy number counts at these wavelengths are therefore a powerful
probe of the relationship between stellar mass and galaxy evolution.
By measuring these number counts, we can constrain models of how
galaxies evolve and thus improve our understanding of the underlying physics.

The Cosmic Infra-Red Background (CIRB) is formed at least in part by
the light from faint, source-confused galaxies.  Measurements of the
faint galaxy number count distributions therefore provide us with
information about the CIRB, and vice versa.  In recent years,
a number of direct measurements have been made of the CIRB 
\citep[see e.g.][]{DwekArendt-98, Gorjian-00, WrightReese-00,
  Matsumoto-05}.  Comparison between such estimates and those arising
from determination of galaxy number counts can substantially enhance
our understanding of the nature of the CIRB. 

One potential limitation to any astronomical observation is that of
source confusion \citep[see e.g.][]{Scheuer-57, Scheuer-74, Condon-74}.
For any finite-resolution observations, sources sufficiently close to
one another on the sky will seem to overlap, or confuse one another.  This
confusion limits the accuracy to which the flux and position of
individual sources can be measured, hence being a source of noise for
the observation.  

This limitation can be therefore particularly acute for IR
observations.  The longer wavelength (relative to optical), combined
with the necessity for space-based instrumentation (with the implied
smaller telescope aperture) tend to lead to relatively poor
resolution, relative to that of ground-based optical observations.  
For deep Spitzer observations in particular, confusion noise becomes the dominant
noise contribution. The Spitzer GOODS survey
\citep[][]{SpitzerGOODS-2004} in particular
is significantly confusion-limited.  If we are to extract
maximum information from these observations, we must resort therefore to
statistical methods.

The extraction of information from source confusion is not a new
problem in astronomy.  As early as the
1950s, radio astronomers were aware that finite resolution would lead
to confusion noise in their observations.  It was realised, however,
that information about the confusing sources was contained in the
Probability Density Function (PDF) of this confusion noise.  So-called fluctuation
analyses, fitting model source distribution PDFs to that of the data, 
have been used in a number of different branches of astronomy over
the years \citep[see e.g.][]{radioPDAnalysis-82, XrayPDAnalysis-93,
XrayPDAnalysis-94, ISOpdAnalysis-97}.
allowing statistical information about the confusing population of
galaxies to be extracted when the individual galaxies cannot be
resolved.  

These fits have typically relied on chi-squared statistics to
fit model PDFs to the data, with a Gaussian form assumed for
the likelihood function in order to estimate errors.  Modern statistics
however, can do more.  Markov Chain Monte Carlo (MCMC) sampling \citep[see e.g.][]{GilksMCMC-book}
gives us a powerful tool for mapping the posterior distribution.  By fully
mapping the posterior, we can estimate both the maximum likelihood
point and the uncertainties on our measurement, without needing to
make any assumptions as the Gaussianity of the error distribution of
the parameters under consideration.

In this paper, we present a fluctuation analysis of the Spitzer GOODS survey,
in order to constrain the sub-confusion Spitzer galaxy number counts at 
3.6, 4.5, 5.8 and 8 microns. In addition to a maximum
likelihood fit, we use MCMC
sampling to determine the uncertainties on our parameter estimates.

The contents of this paper are therefore as follows.
\begin{description}
\item{In {\tt section \ref{section:SpitzerGOODSdata}}, we detail the Spitzer GOODS survey.}
\item{In {\tt section \ref{section:methods}}, we discuss fluctuation analysis, both in general terms and
in detail specific to the analysis presented in this paper.}
\item{In {\tt section \ref{section:analysisResults}}, we present the results of our analyses.}
\item{Finally, our conclusions are presented in section {\tt \ref{section:conclusions}}.}
\end{description}
\section[]{The Spitzer GOODS data} \label{section:SpitzerGOODSdata}
The Great Observatories Origins Deep Survey (GOODS) is a series of
observing programs that are creating a public multi-wavelength
data set for studying galaxy formation and evolution.  It comprises
observations from the Spitzer, Hubble, Chandra and XMM-Newton space
observatories, with extensive follow-up from ground-based facilities.

The Spitzer GOODS observing programme \citep[][]{SpitzerGOODS-2004}
comprises deep observations in the vicinity of the Hubble
Deep Field North (HDF-N) and the Chandra Deep Field South (CDF-S).
The observations cover a total of approximately 300 arcmin$^2$, using
all four IRAC bands \citep[][]{SpitzerIRAC-2004}
(3.6, 4.5, 5.8, 8 microns), with 24 micron MIPS observations also planned.

In this paper, we analyse the Spitzer GOODS super-deep epoch one data
release for the HDF-N.  For each of the four IRAC bands, we analyse
the full image data from this field.  This gives us ($9.2\times10^5,
9.5\times10^5, 9.2\times10^5, 9.5\times10^5$) 'good' (i.e. unflagged)
pixels in the bands.  As our measurement is a statistical inference,
we adopt a conservative flagging regime, retaining only pixels
containing 50\% or more of the model exposure time.  This ensures we
use the best signal:noise data, as well as those pixels with the
best-defined noise estimates.

The data release includes rms estimates of the shot noise component
due to the sky background and instrument noise.  We use these as our
overall instrumental noise estimate, assuming that the noise for each
pixel is Gaussian and uncorrelated with that of its neighbours (true by
construction here, due to use of a point-kernel drizzle algorithm in
construction of the maps).  We test the impact of any inaccuracy in these
assumptions by varying the noise estimates by 20\%; there is
negligible impact on the final results, so we conclude that this noise
treatment is sufficient.

One other issue is that of the effect of flat-fielding.  It is known
(Surace, private communication) that this process introduces a low
level of residual fluctuation into IRAC images such as the ones
analysed in this paper.  These fluctuations are correlated on short
angular scales (of order that of the PSF), but should be uncorrelated
with the sky; as our analysis considers only the one-point statistics
of the data, this effect merely represents an additional source of noise
in the data (albeit one for which we have not explicitly accounted).  However,
provided the resultant noise is Gaussian in nature, our above noise
tests show that our analysis is robust to their effects.  We note that
this effect is known to be most significant in the shorter wavelength
IRAC bands.

The data release images have already had a background subtraction run
on them.  In addition, we also perform what amounts to a basic
background subtraction by subtracting the overall median from both the
data and model pixels, in order to properly match them.

We also identify and mask out bright sources in the images.  We do
this by simply identifying pixels above a flux threshold, then masking
out a surrounding region sufficient to remove the entire source.  Due
to the large number of available pixels, we conservatively
overestimate the number of pixels that need to be masked, in order to
avoid bias in our results.

The HDF-N field is an excellent data set for fluctuation analyses in
that the statistical errors are likely to be the sub-dominant part of
the uncertainty in the final results.  Systematic effects such as
absolute calibration uncertainty and model mis-matching will be more
significant.  It is therefore unnecessary to also analyse the southern
field.

\section[]{Methods} \label{section:methods}

\begin{figure*}
\begin{minipage}{150mm}
\epsfig{file=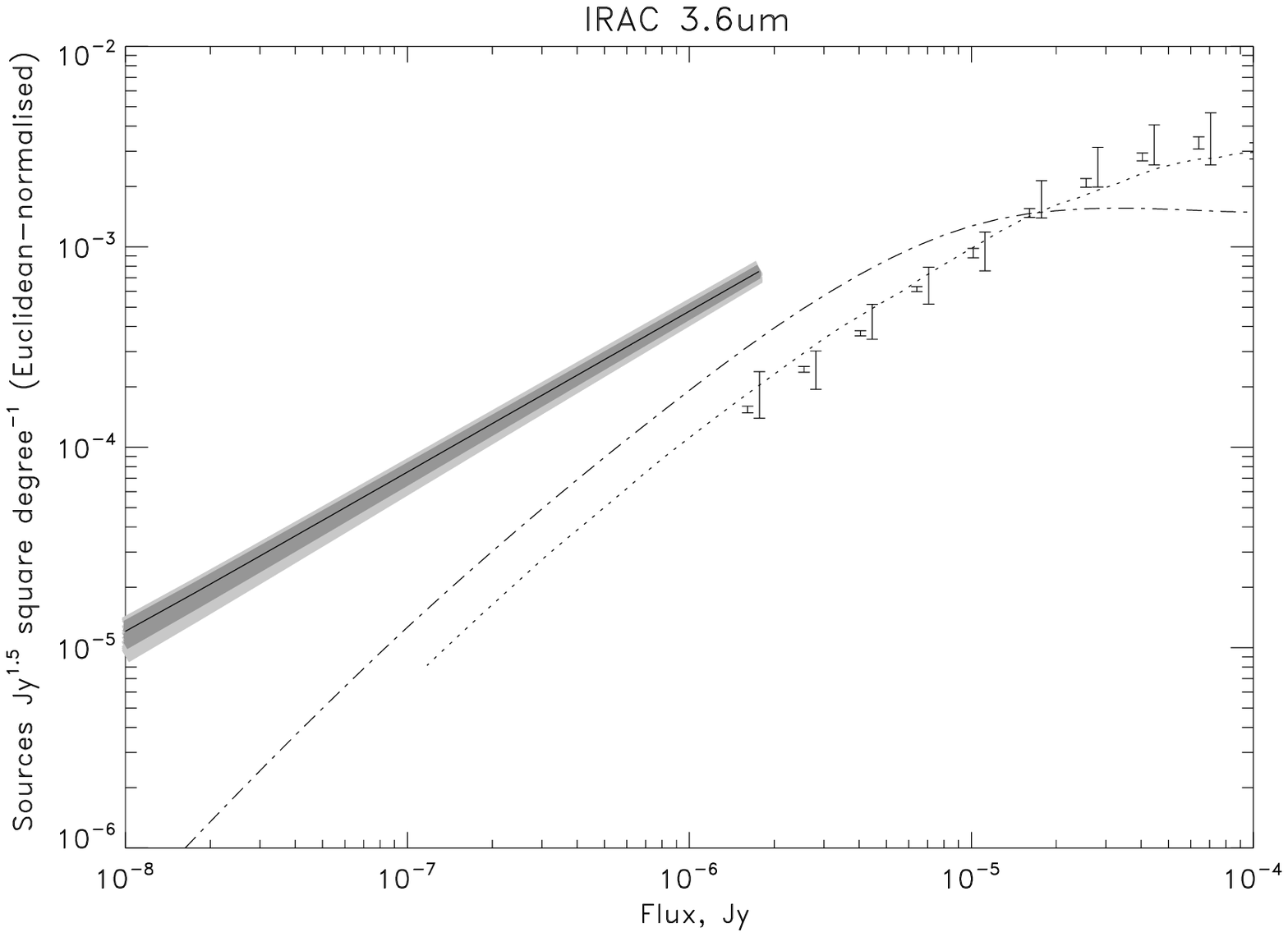  , angle=0, width=8.5cm}
\epsfig{file=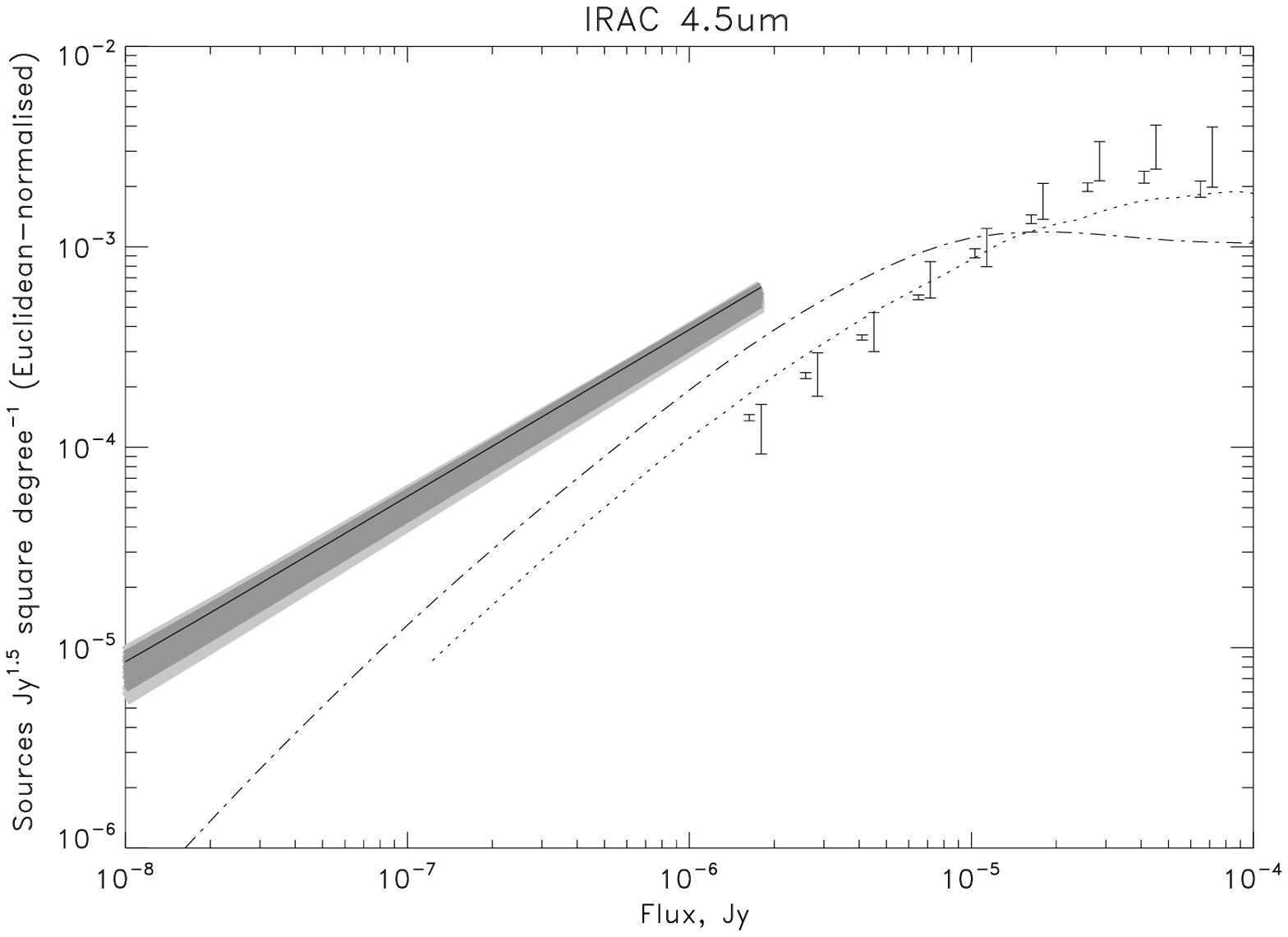  , angle=0, width=8.5cm}
\epsfig{file=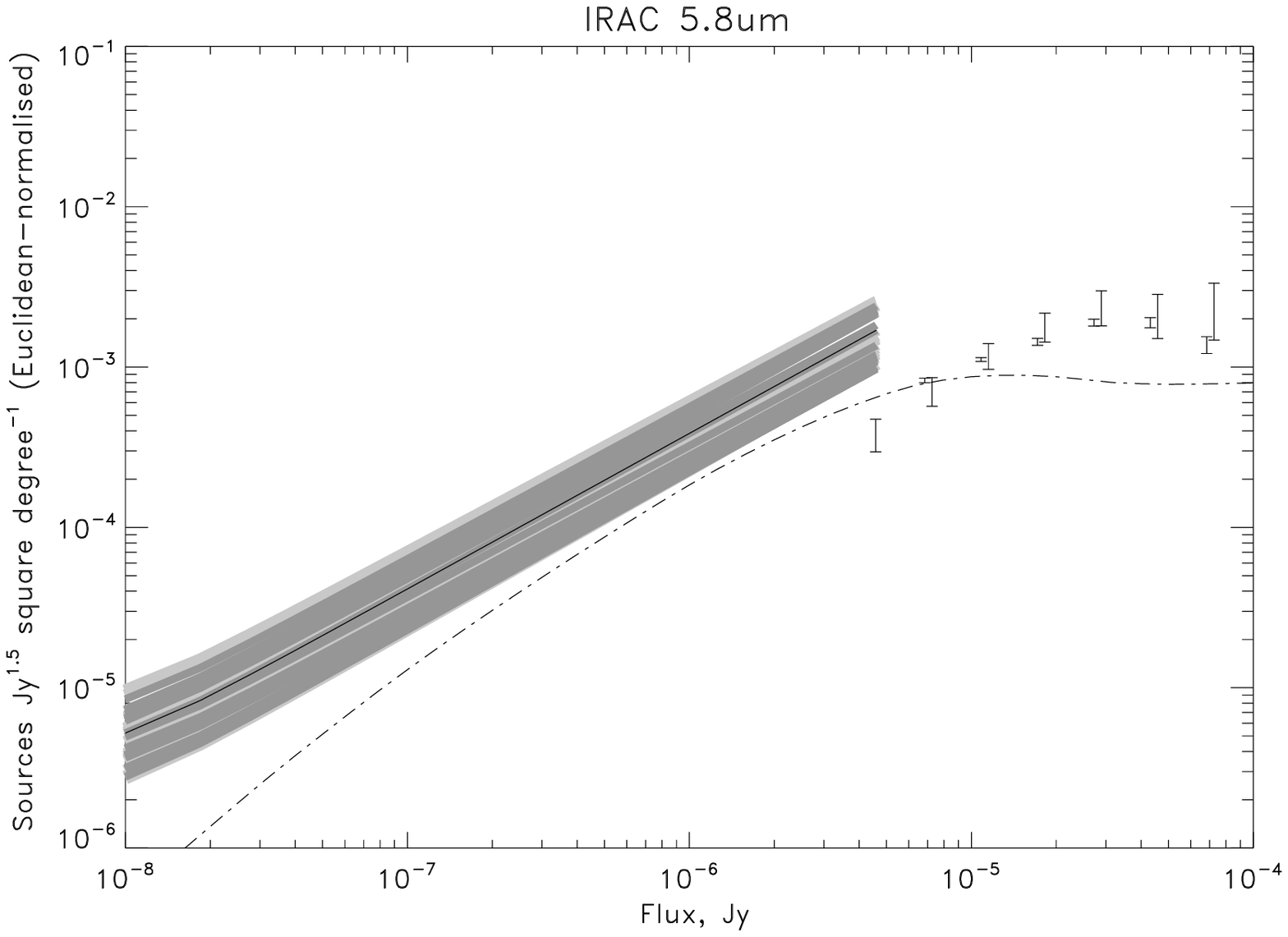  , angle=0, width=8.5cm}
\epsfig{file=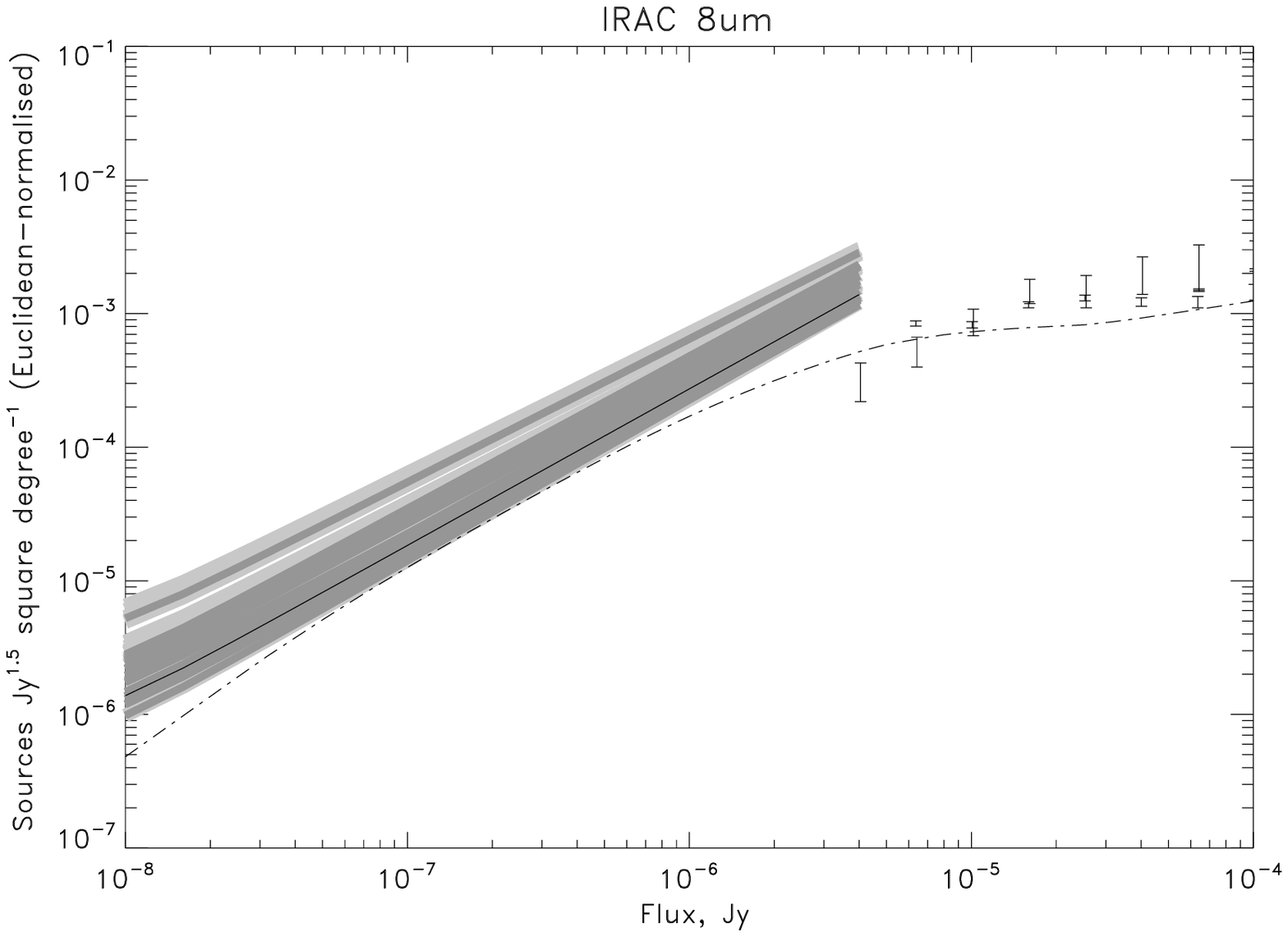  , angle=0, width=8.5cm}
\caption{Plots of the fitted model number counts.  In each case, the black line
  denotes the best-fit model, the grey regions are generated by
  plotting a sample of the MCMC models with the highest 68\% and 95\%
  posterior probability values, respectively.  Note that all models
  are Euclidean normalised by multiplying by $flux^{2.5}$.  The
  plotted points are actual number counts, measured by
  \citet[][]{Fazio-04}.  Also plotted are model number count
  predictions due to \citet[][]{Pearson-05} (dot-dash line) and
  \citet[][]{Franceschini-05} (dotted line).
  }
\label{fig:numberCountFits}
\end{minipage}
\end{figure*}

Our aim is to extract statistical information about the
nature of the faint galaxy population contributing to the confusion noise in
the Spitzer GOODS survey.  To achieve this, we will use a fluctuation
analysis.  

Fluctuation analyses uses the PDF of the data image pixels to constrain a
parameterised model of the faint source number counts. The model is fitted to
the data by comparing the PDF of the data image pixels with the
corresponding PDF predicted for the chosen
model.  Various model-fitting techniques (in this case Markov-Chain Monte Carlo
sampling) can then be used to determine the most likely set of
parameter values for the model, given the data, as well as to
determine the uncertainties on these estimates.

\subsection{PDF estimation}
In order to carry out a fluctuation analysis, we need a method of
estimating the PDF of our data.  The most straightforward way of doing
this is to take a histogram of the image pixels, and this has several advantages.

Histograms are straightforward to implement and also to 
determine a reasonably 'optimal' bin width \citep[see
  e.g.][]{Scott_densityEstimation}.  It is also easy to construct a
histogram such that the bins are essentially uncorrelated from one
another, giving the individual bins Poissonian errors.  Furthermore,
for large numbers of pixels-per-bin, the errors tend towards
Gaussianity, a feature that simplifies the calculation of likelihoods.

We note that there are a number of other possible choices of 
PDF estimators.  In particular, fixed and adaptive kernel methods
\citep[see e.g.][]{Vio-94} have become increasingly widely used in recent
years.  The superior performance of these methods comes at the price
(in this context) of a more complicated error structure and
greater computational requirements.  We will therefore not consider
them in this work.

\subsection{Generation of model PDFs}
Having estimated the PDF of our data, we need equivalent
model PDFs to fit to it.  In previous work, these have typically been
generated from analytic solutions \citep[see e.g.][]{Condon-74}.  In
this paper, however, we will adopt a slightly different approach,
designed to more explicitly reflect the effects for which we must
account.  

Our method of determining model PDFs is essentially a Monte Carlo
method, simulating a number of skies in order to find the required
PDF.  However, in order to speed up the calculation, we implement
numerically-solved analytic solutions in order to determine the
distribution of fluxes arising from any pixel on the sky (before the
effect of the instrument point spread function), as well as the effect
of instrumental noise.

Due to the nature of fluctuation analysis, we pay particular attention
to dealing with the instrumental noise.  The data used in this paper
include pixel-by-pixel estimates of the noise variance.  To properly
account for variations in noise across the map, we generate the
overall noise PDF by summing the Gaussian noise PDFs (characterised by
said variances) for each pixel (and re-normalising).  The noiseless
model PDF is then convolved by this noise PDF and the resulting model
PDF re-binned to match the histogram bins of the data PDF.

To test the robustness of our analyses to noise estimation, we ran
test analyses in which we increased or decreased the noise variances by
20\%.  In all cases, the results were essentially unchanged, an
outcome that is reasonable given that source confusion is the dominant
effect in these observations.

We note that this method is all carried out on pixels of twice
the resolution of the data we will be analysing.  This matches the
highest resolution publicly available point spread function for the
IRAC instrument, allowing us to use all available information about
the point spread function.  The model resolution is halved by simple bin
averaging \emph{after} the point spread function convolution, but
\emph{before} convolution with the noise PDF.

\subsection{Model-fitting, mapping the posterior probability distribution}
For any analysis such as this, we are aiming to determine the most
likely set of model parameters, given our data (and choice of model).
From fundamental probability theory, we have Bayes theorem:

\begin{equation}
P(\theta|D,H)=\frac{P(D|\theta,H)P(\theta,H)}{P(D|H)}\,,
\label{BayesTheorem}
\end{equation}

Where $P(\theta|D,H)$ is the posterior probability of the model
parameters ($\theta$), given the data ($D$) and a hypothesis ($H$).
$P(D|\theta,H)$ is the likelihood of the data given a set of model
parameters, $P(\theta,H)$ represents any prior
knowledge we may have and $P(D|H)$ is the Bayesian Evidence.

So, the posterior is the thing we want.  Assuming uncorrelated,
Gaussian errors on the PDF histogram bins, the (unnormalised)
likelihood is given by the following equation.

\begin{equation}
L=exp(-{Q^2\over2})
\end{equation}

where $Q^2$ is given by

\begin{equation}
Q^2=\sum_{i=1}^{N_{bins}}\left({{d_i-m_i}\over\sigma_i}\right)^2
\end{equation}

and $d_i$ is the number of data in the $i^{th}$ histogram bin, $m_i$
is the corresponding model value and $\sigma_i$ is the standard
deviation associated with that histogram bin.

For this analysis, we choose uninformative flat priors, so that we are
making minimal assumptions.

For the analyses presented in this paper, we will ignore the
normalising Bayesian evidence term.  However, we note that future work
in this will undoubtedly benefit from using the evidence to compare
the relative likelihood of different number count models being good
descriptions of the data.

The information extracted by such an analysis is to be found in
the posterior probability function.  This function tells us, given our
data (and any prior knowledge), the likely range of values of the
model parameters. We wish therefore to map this function, in order to produce the
results of our analysis.

The analysis methods we have described in the previous subsections
allow us to determine the (unnormalised) posterior value for a given
point in parameter space.  By using some suitably efficient method of
function mapping, we can hence obtain our results.

We will use Markov Chain Monte Carlo (MCMC) sampling \citep[see
  e.g.][]{GilksMCMC-book} to do this.  MCMC sampling is an efficient
way of drawing a set of samples from the posterior probability
distribution.  It has a number of advantages over many more basic
methods.  Firstly, it returns not only the most likely
solution, but also estimates of the uncertainties on all estimated
parameters.  Secondly, it estimates these uncertainties without
needing to make any assumptions as to the functional form of the
likelihood (such as are required by methods such as Fisher matrix analysis).  The
principal downside is that such an analysis can be computationally
very expensive; typically $5 \times 10^4$ to $10^5$ likelihood calculations
are needed, several orders of magnitude more than most function maximisers.

In each of the four analyses we carry out, we use at least five MCMC
chains, each 10,200 samples in length.  We remove the first 2,000
samples from each case as they represent a 'burn-in' phase, where the
chains are moving towards the region of highest likelihood.  This
leaves us with over 40,000 samples for each analysis.  

Ideally, we would want more samples than this (a typical rule-of-thumb
for this type of analysis is $10^5$ or more samples), however we were
limited in this work by the amount of time taken to compute the
likelihood values; the analysis for each band took 3-4 weeks on a 3~GHz,
dual-processor Linux box.

\subsection{Choice of models}
In this paper, we will fit the usual class of power law
models for the differential source number counts, defined by the
following equation.

\begin{equation}
{dN \over dS} = \left\{
\begin{array}{ll}
N_o S^{-\delta} &  (S>S_{cut})\\
0             &  (S<S_{cut})\\
\end{array}
\right.
\end{equation}

The fitted parameters are $\delta$, $S_{cut}$ and $A$, where $A$ is
given by the following equation.

\begin{equation}
A = \int_{0}^{S_{max}} {dN \over dS} dS
\end{equation}

$S_{max}$ is chosen to exclude small numbers of bright sources
at a level where the model will no longer be a fit to the observed
number counts.  We note that in each analysis we have removed regions
of the data containing flux peaks bright enough to be due to sources
brighter than the corresponding value of $S_{max}$.

The sources are assumed to be Poisson-distributed on the sky (i.e. we
assume they are not clustered).  We note that this model is strictly
phenomenological. This approach, while not physically
motivated, does produce constraints on the likely number count
distribution of faint sources in the fields being analysed.

\section[]{Analysis and results} \label{section:analysisResults}

\begin{table*}
\begin{minipage}{150mm}
\caption{The results of fitting a one-power-law number count model to
  the data.  For each band, the best-fit log-likelihood, number of
  histogram bins, best-fit mean source density, power law index, flux
  cut-off are given, along with the 95\% confidence limits (determined
  from the 1D marginalised likelihood functions(see figure
  \ref{fig:1dLikelihoods}).  Reduced chi-squared values are also
  given.}
\label{table:onePowerLaw}
\begin{tabular}{lccccccc}
\hline
Band & Log(likelihood) & bins & source density ($10^{9}$ per sq. deg.) & index &
cut-off ($10^{-12}$ Jy) & reduced chi-squared \\
\hline
3.6 microns & -162.7  & 98 & 1.62$^{+0.62}_{-0.62}$ & 1.70$^{+0.01}_{-0.05}$ & 15.0$^{+8.5}_{-8.5}$  & 3.43 \\
4.5 microns & -41.0   & 99 & 2.14$^{+2.82}_{-1.24}$ & 1.67$^{+0.02}_{-0.04}$ & 4.01$^{+4.75}_{-2.70}$ & 0.86 \\
5.8 microns & -54.1   & 98 & 1.44$^{+1.82}_{-0.85}$ & 1.53$^{+0.05}_{-0.03}$ & 0.24$^{+1.37}_{-0.22}$ & 1.14 \\
8.0 microns & -54.3   & 99 & 0.09$^{+1.17}_{-0.03}$ & 1.33$^{+0.14}_{-0.02}$ & 0.54$^{+0.75}_{-0.53}$ & 1.13 \\
\hline
\end{tabular}
\end{minipage}
\end{table*}

\begin{table*}
\begin{minipage}{150mm}
\caption{The integrated background light due to galaxy emission, for
  each of the four IRAC bands.  These values are estimated in two
  parts.  \emph{Faint} is found analytically by integrating the
  best-fit model from our fluctuation analysis. \emph{Bright} is the
  integrated light due to sources above the upper limit of our models
  and are taken from \citet[][]{Fazio-04}.  The total is the sum of
  the previous two values.  All errors shown are 95\% confidence
  limits.  \emph{Faint} errors are determined from the 1D marginalised
  likelihood functions (see figure
  \ref{fig:backgroundHistogram}). \emph{Bright} errors are not given in
  \citet[][]{Fazio-04}, so the total error is found by scaling those for
  \emph{Faint}.}
\label{table:integratedBackground}
\begin{tabular}{lccccccc}
\hline
Band & faint ($10^{-9} w m^{-2} sr^{-1}$) & bright ($10^{-9} w m^{-2}
sr^{-1}$) & total background ($10^{-9} w m^{-2} sr^{-1}$) \\
\hline
3.6 microns & 5.23$^{+0.31}_{-0.96}$ & 5.4 & 10.6$^{+0.63}_{-1.95}$ \\
4.5 microns & 2.85$^{+0.43}_{-0.48}$ & 3.5 & 6.4$^{+0.97}_{-1.08}$  \\
5.8 microns & 1.79$^{+0.56}_{-1.18}$ & 3.6 & 5.4$^{+1.69}_{-3.56}$   \\
8.0 microns & 1.23$^{+0.77}_{-0.74}$ & 2.6 & 3.8$^{+2.38}_{-2.29}$  \\
\hline
\end{tabular}
\end{minipage}
\end{table*}

Figure \ref{fig:numberCountFits} shows the results of our analyses.
In each case, the solid black line denotes the best-fit model (whose
parameters are given in Table \ref{table:onePowerLaw}).  The shaded
regions give the loci of 68\% and 95\% confidence regions, defined by
taking the corresponding percentage of MCMC samples with the highest
posterior probability values.

On each plot, the number counts determined for the four IRAC bands by
\citet[][]{Fazio-04} are plotted.  The values plotted are those
estimated for galaxy counts, with error bars denoting Poisson
uncertainty.  All values have been corrected for incompleteness using
the estimates given in said paper.  The plots also show the galaxy
number count models of \citet[][]{Pearson-05} (dot-dash line).  

\begin{figure}
\epsfig{file=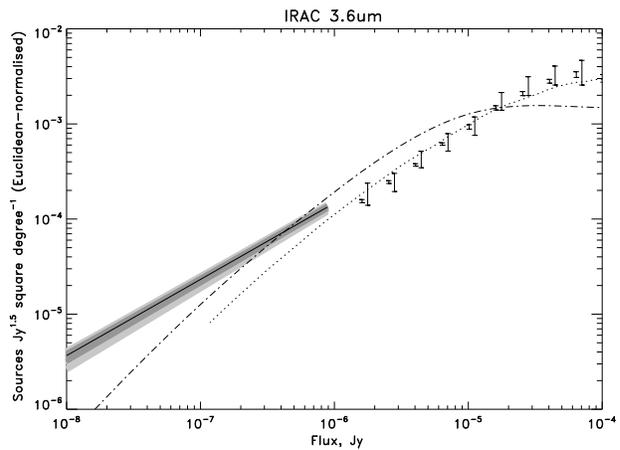  , angle=0, width=8.5cm}
\caption{Plot of the fitted 3.6 micron model number counts, with model
  flux values scaled by $\times 0.5$, to simulate the effect of an
  error in flux calibration.  (Note that this also affects the
  Euclidean normalisation).  This shows that the difference between
  our results and the \citet[][]{Fazio-04} number counts could be
  explained by a 50\% relative difference in the flux calibration of the
  two sets of observations.}
\label{fig:numberCountFluxScaled}
\end{figure}

In Figure \ref{fig:numberCountFluxScaled}, we re-plot the 3.6 micron
number counts with a flux scaling of 0.75, in order to show the effect
of a 25\% (decrease) error in flux calibration.

\begin{figure*}
\begin{minipage}{150mm}
\epsfig{file=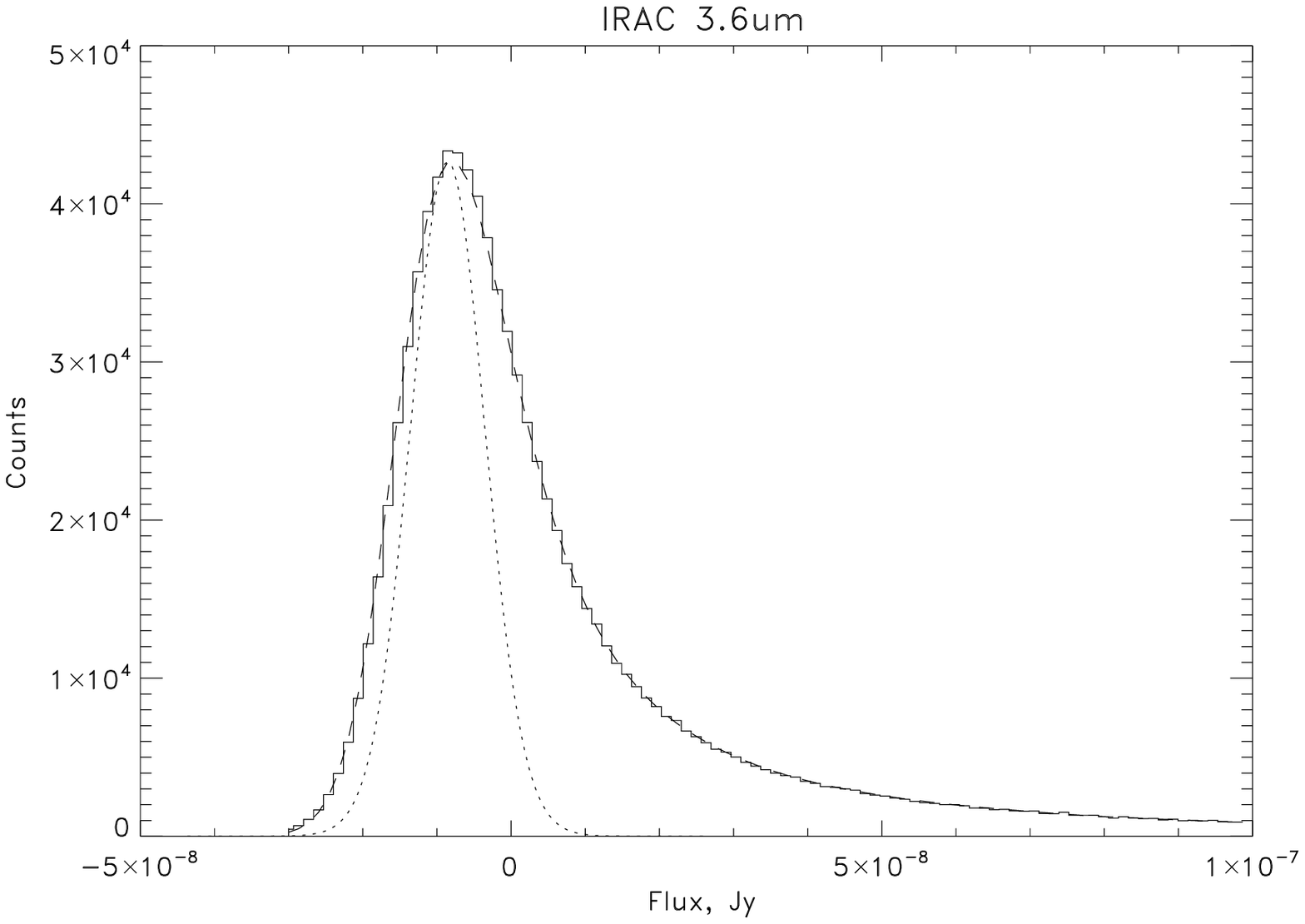 , angle=0, width=8.5cm}
\epsfig{file=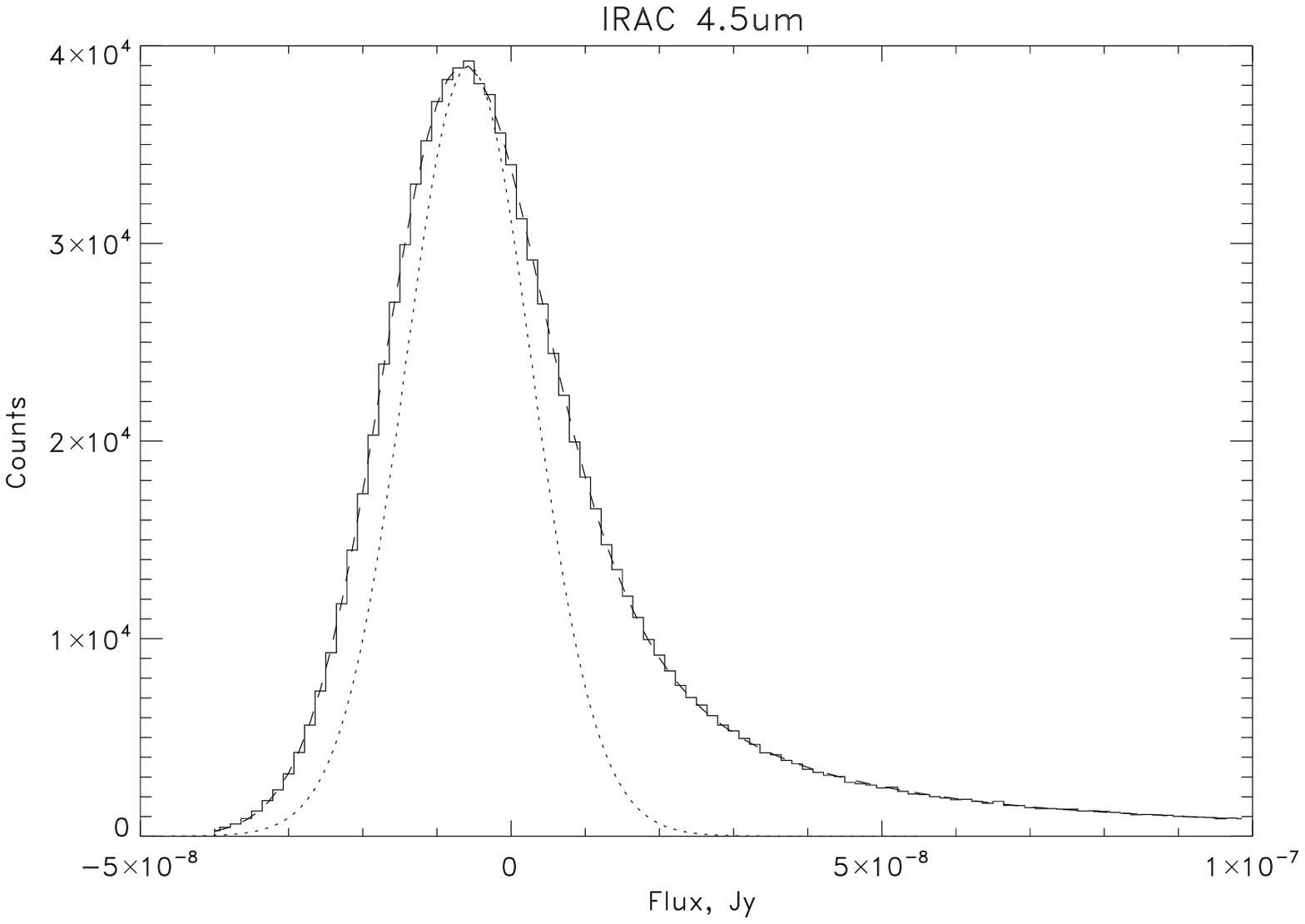  , angle=0, width=8.5cm}
\epsfig{file=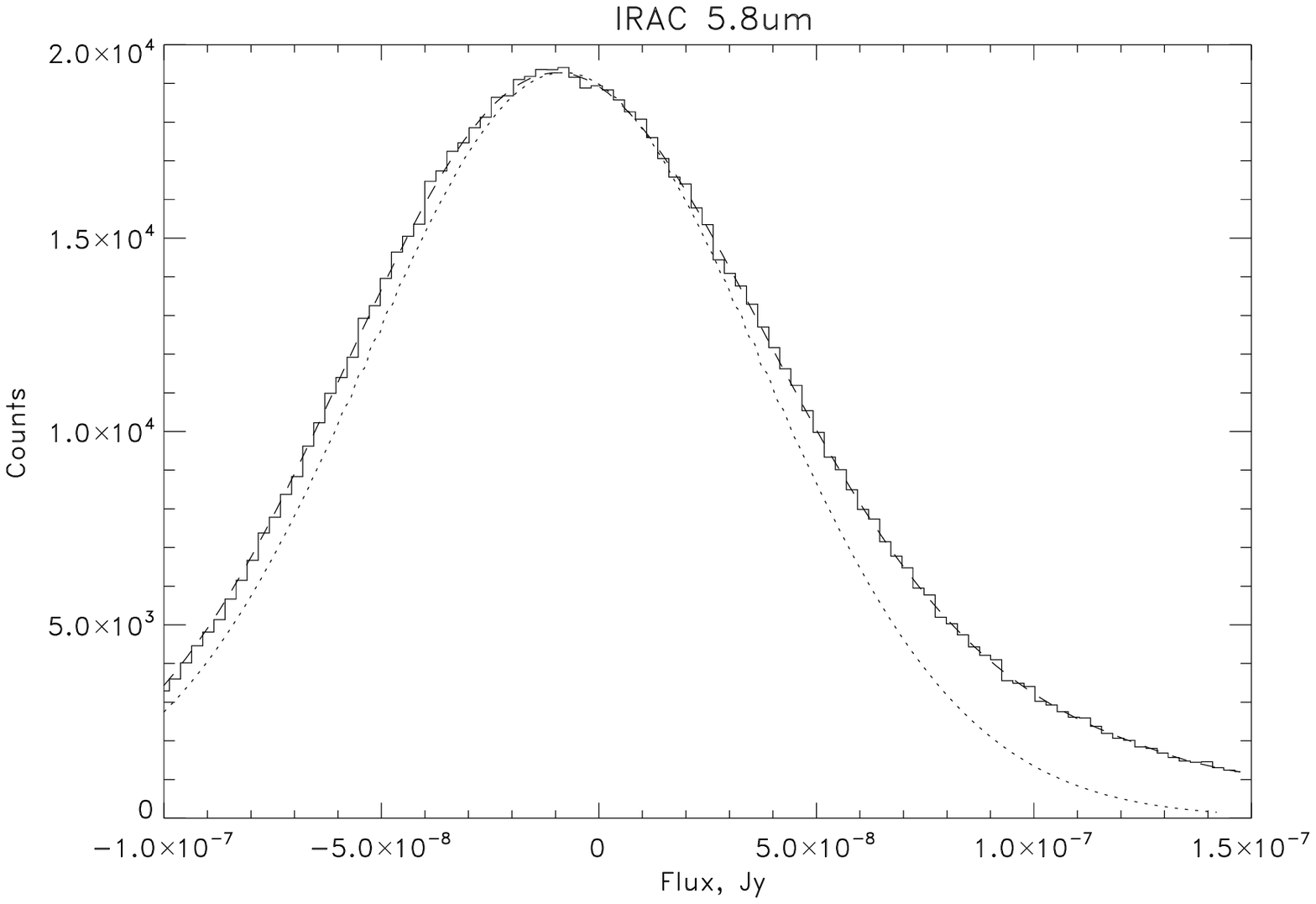  , angle=0, width=8.5cm}
\epsfig{file=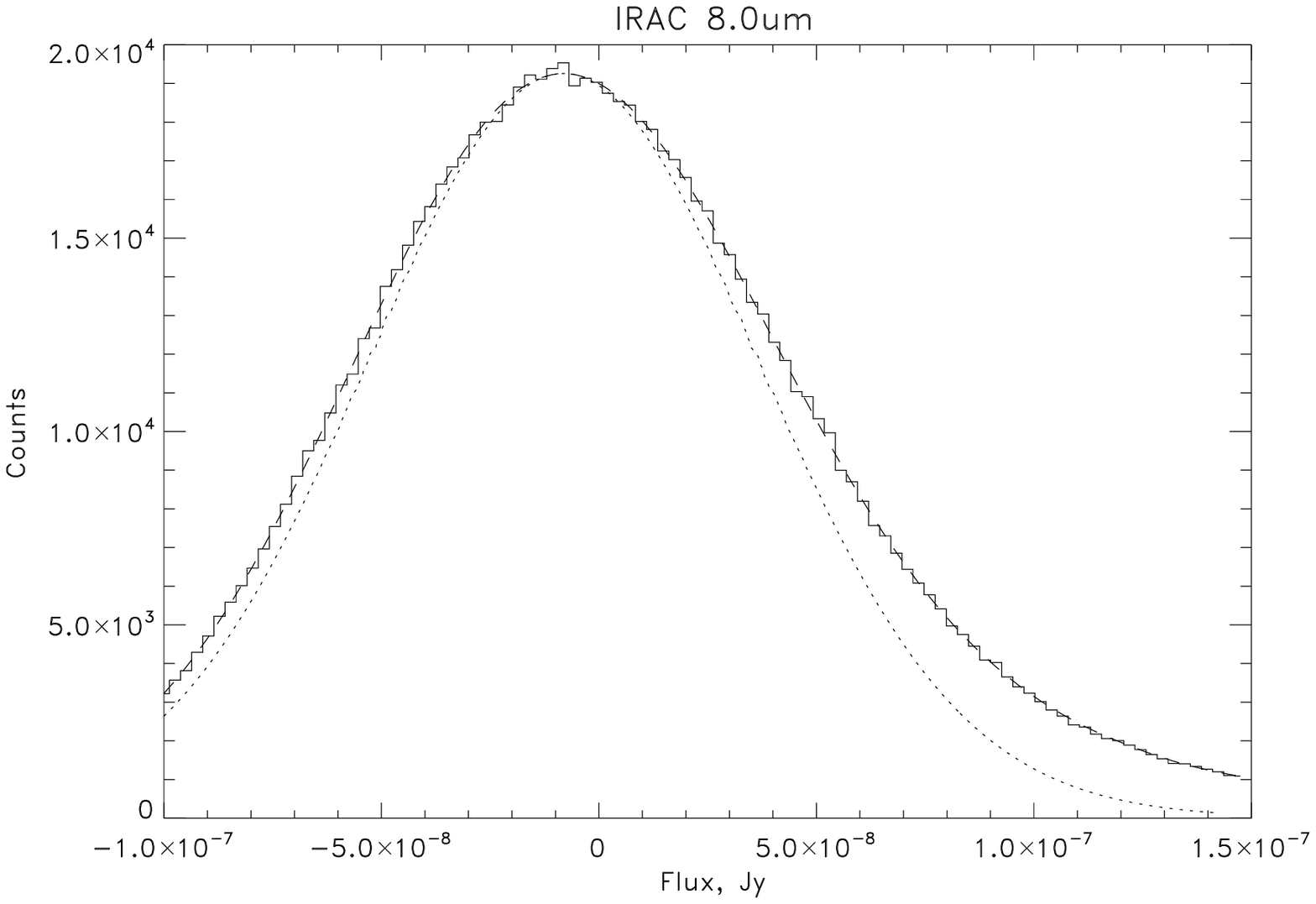  , angle=0, width=8.5cm}
\caption{The pixel flux PDFs for the data (solid line) and the
  best-fit model (dashed line; partially obscured by solid line).
  The data PDF is estimated using a histogram of the image pixels from
  a given band.  The model PDF is calculated using the method
  described in section \ref{section:methods}.  The overall noise PDFs
  are also plotted (dotted line).  For ease of comparison, the noise PDFS are
  appropriately scaled and offset so that their peaks are coincident
  with the data PDFs.}
\label{fig:dataPDFs}
\end{minipage}
\end{figure*}

Figure \ref{fig:dataPDFs} shows the data PDFs estimated for each of
the four IRAC bands (solid lines).  Over-plotted (dashed lines) are the
best-fit model PDFs determined from our analyses.

\begin{figure*}
\begin{minipage}{150mm}
\epsfig{file=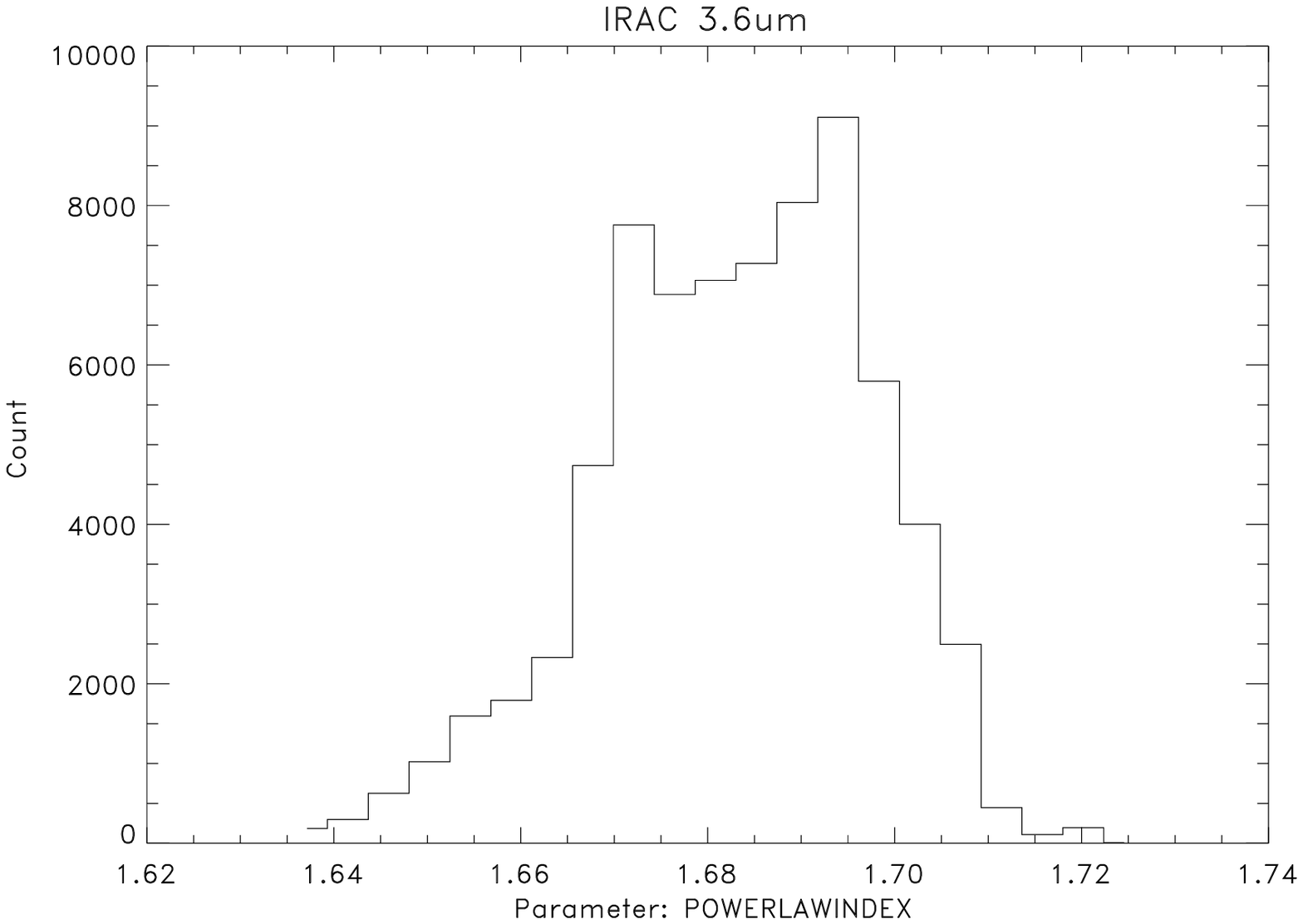 ,     angle=0, width=5cm}
\epsfig{file=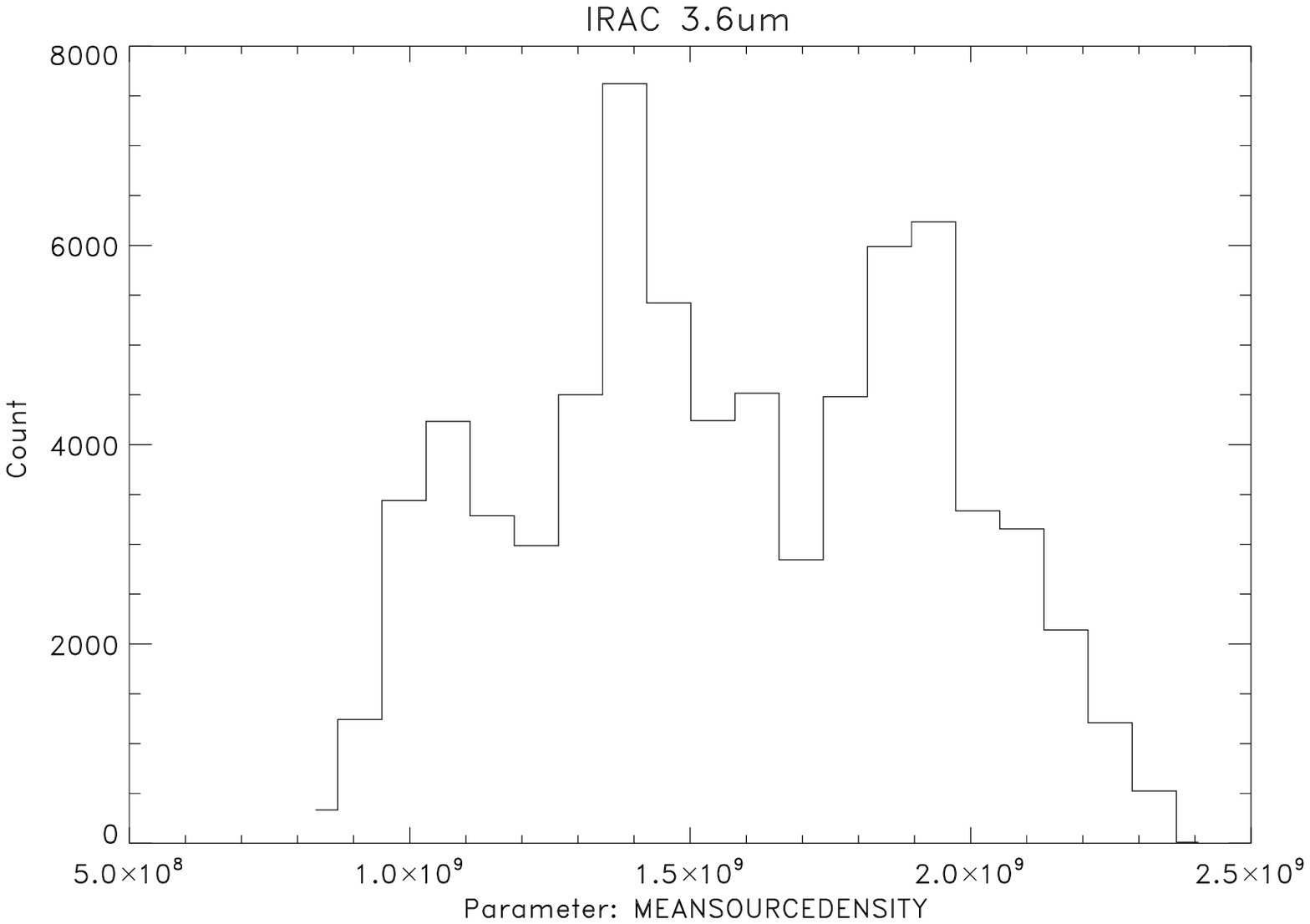 , angle=0, width=5cm}
\epsfig{file=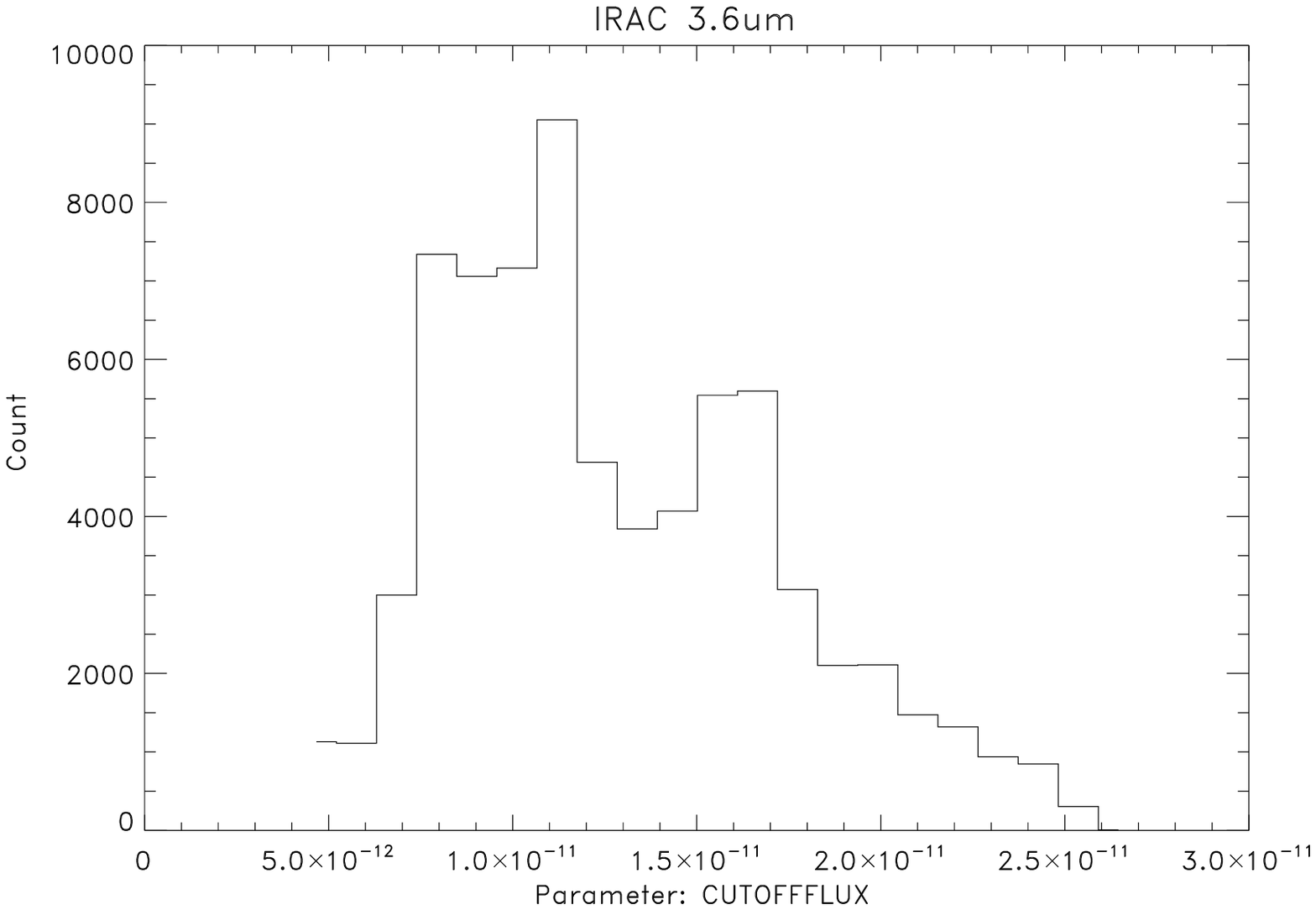 ,        angle=0, width=5cm}
\epsfig{file=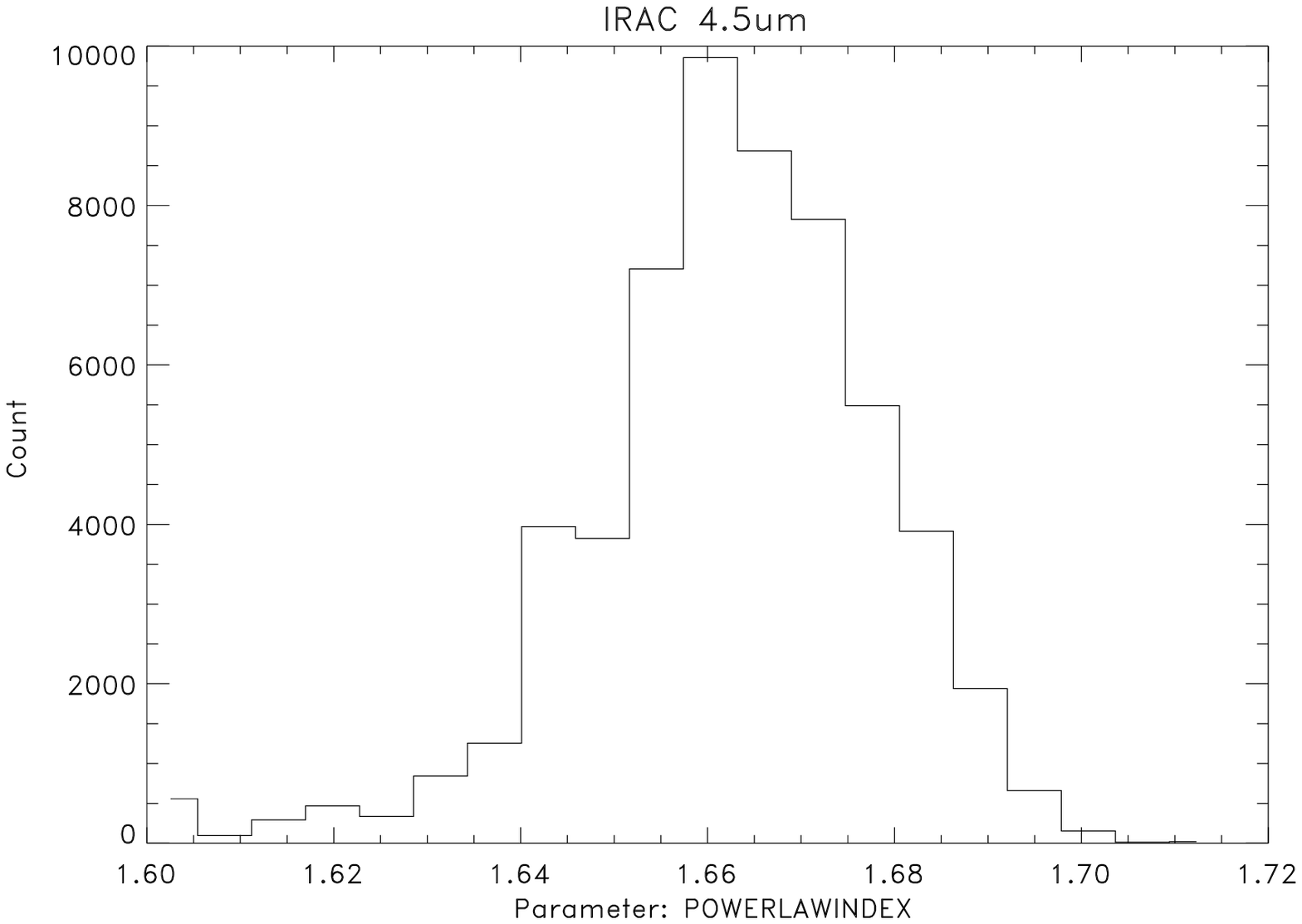 ,     angle=0, width=5cm}
\epsfig{file=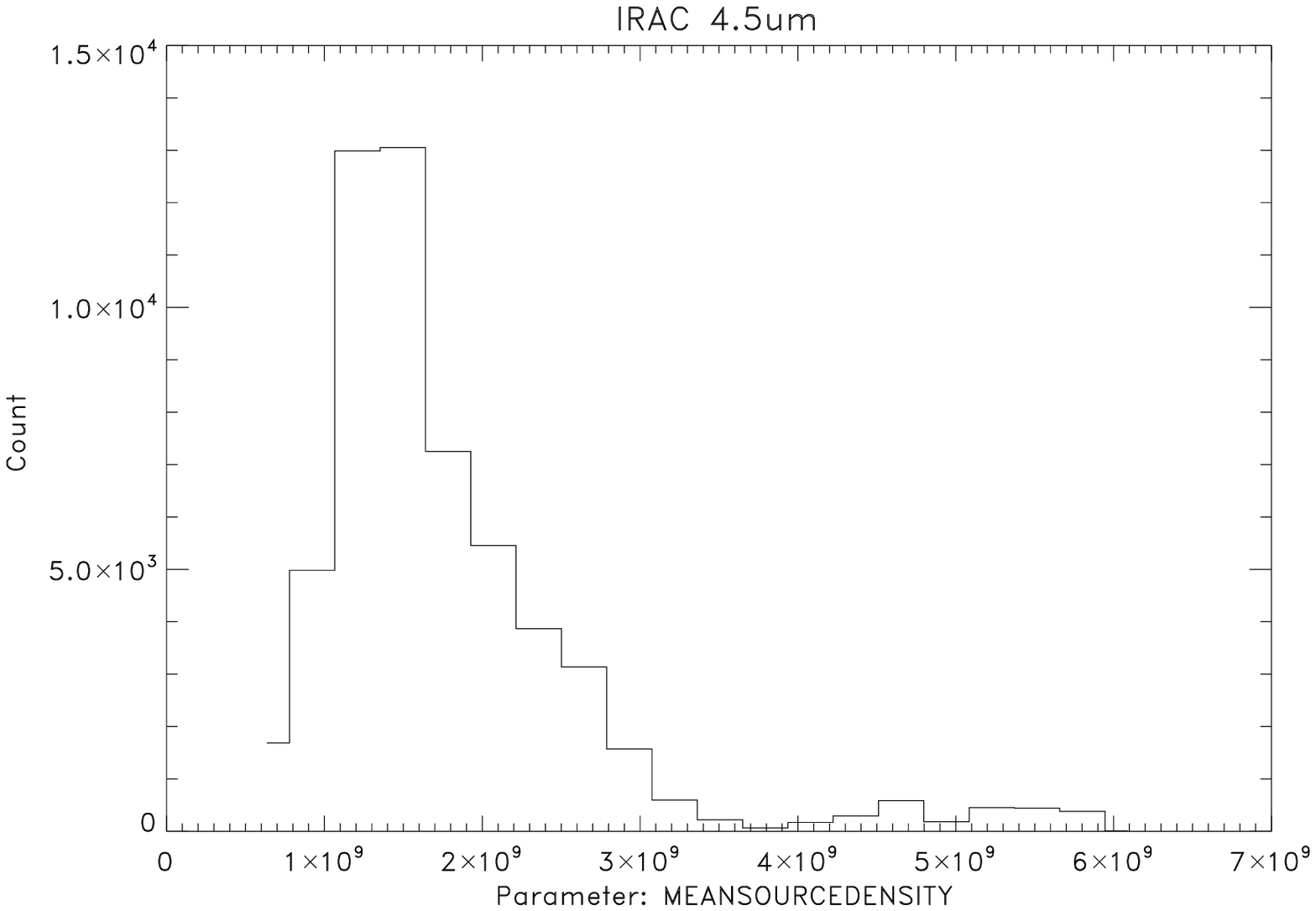 , angle=0, width=5cm}
\epsfig{file=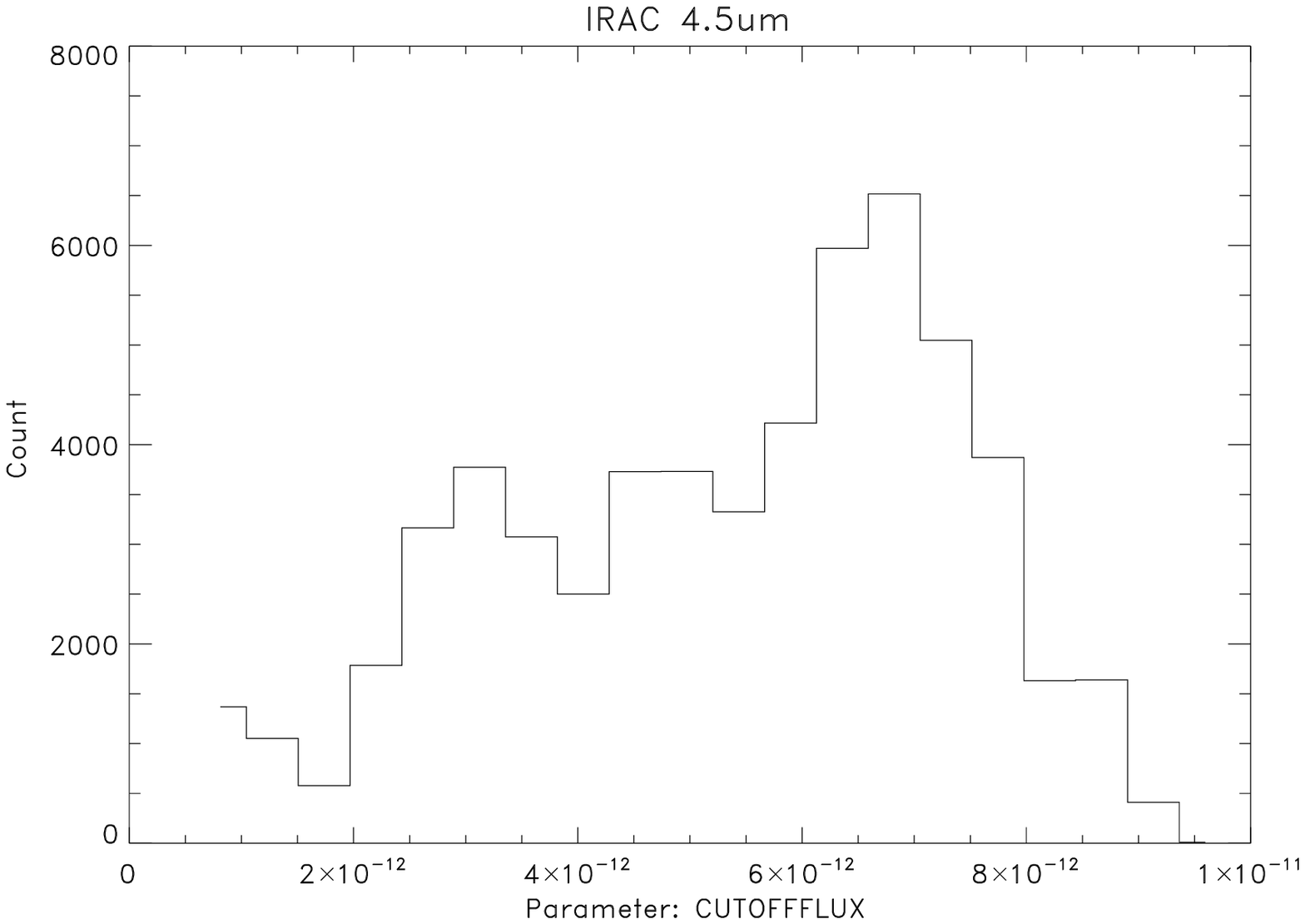 ,        angle=0, width=5cm}
\epsfig{file=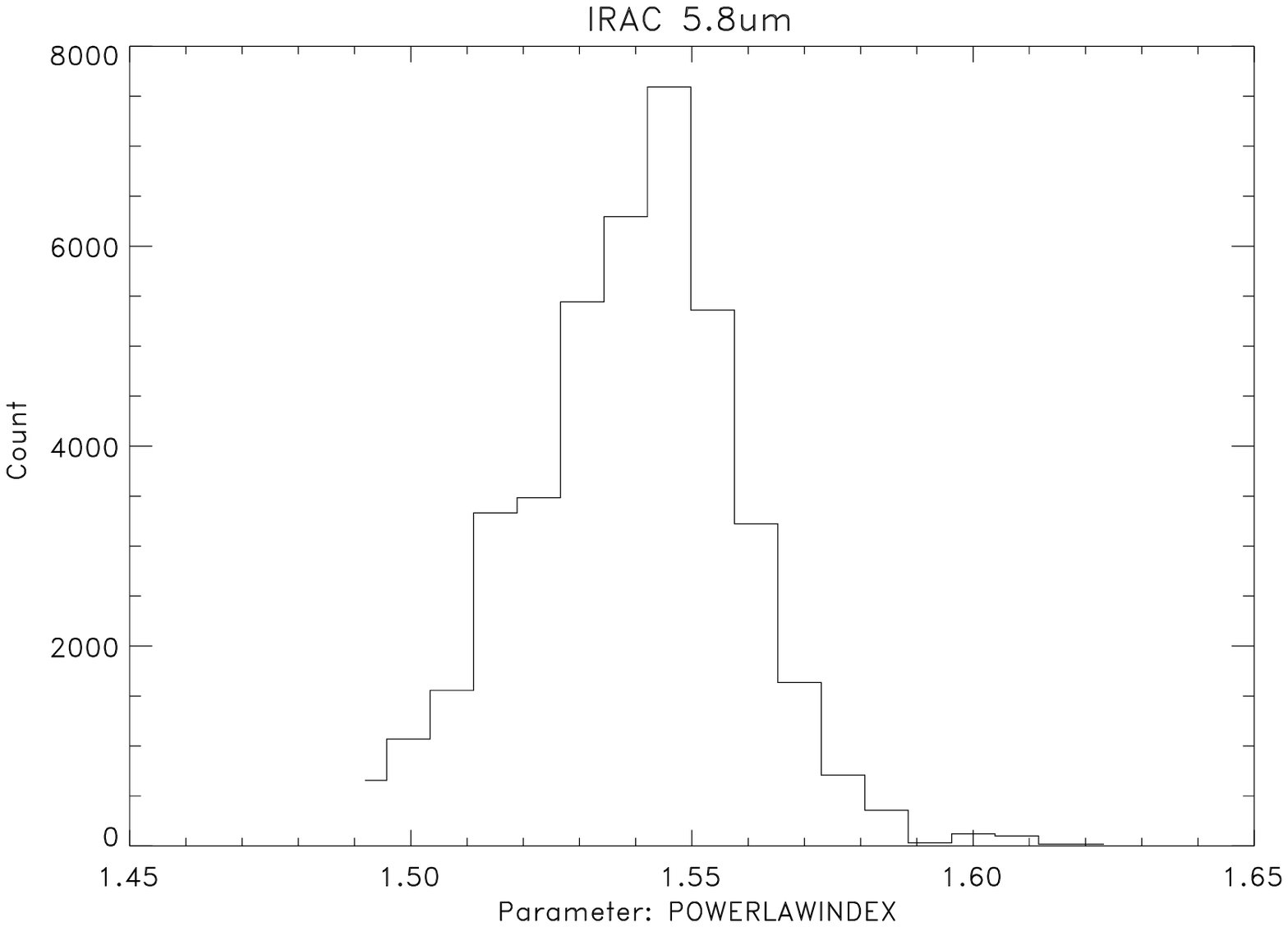 ,     angle=0, width=5cm}
\epsfig{file=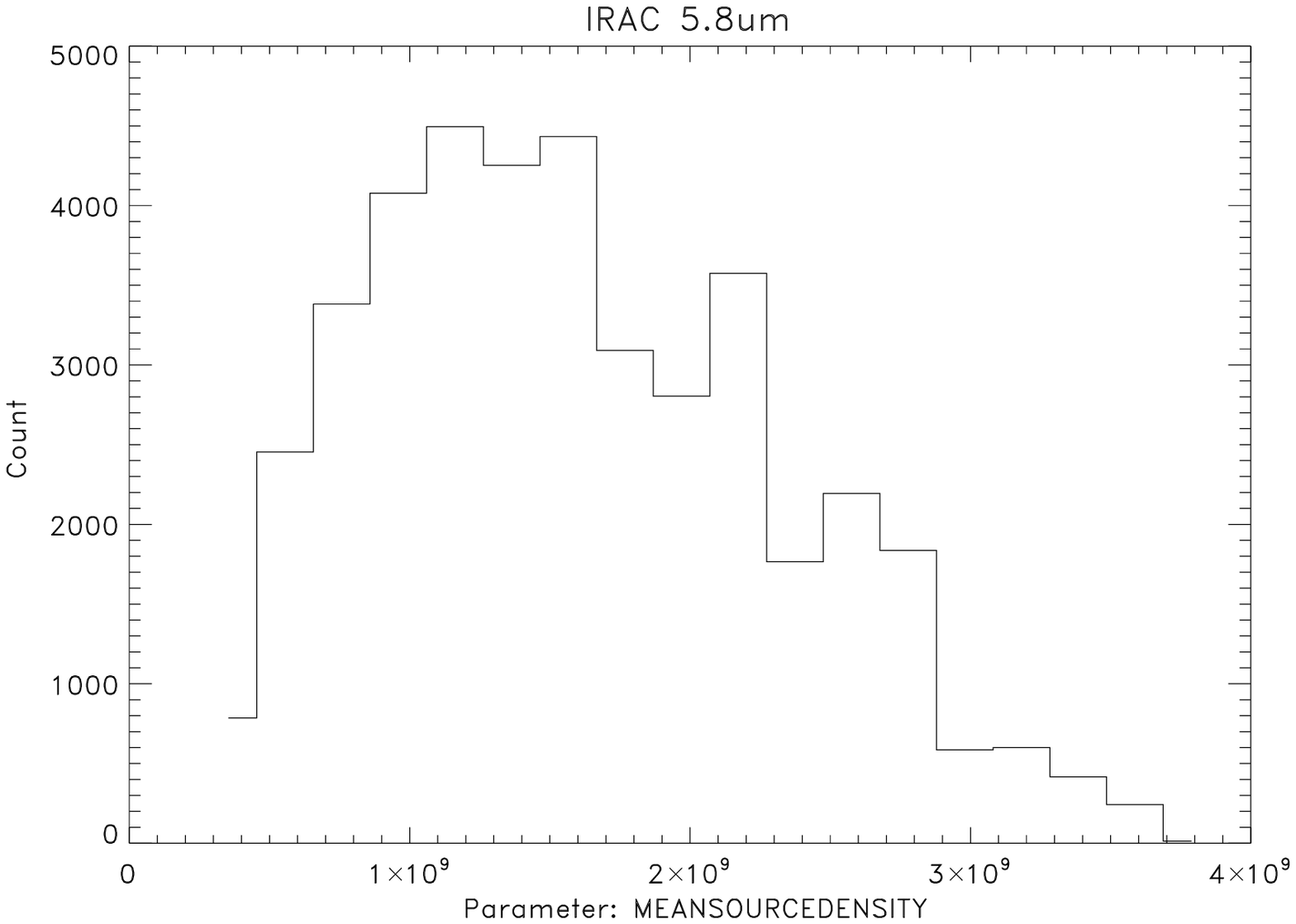 , angle=0, width=5cm}
\epsfig{file=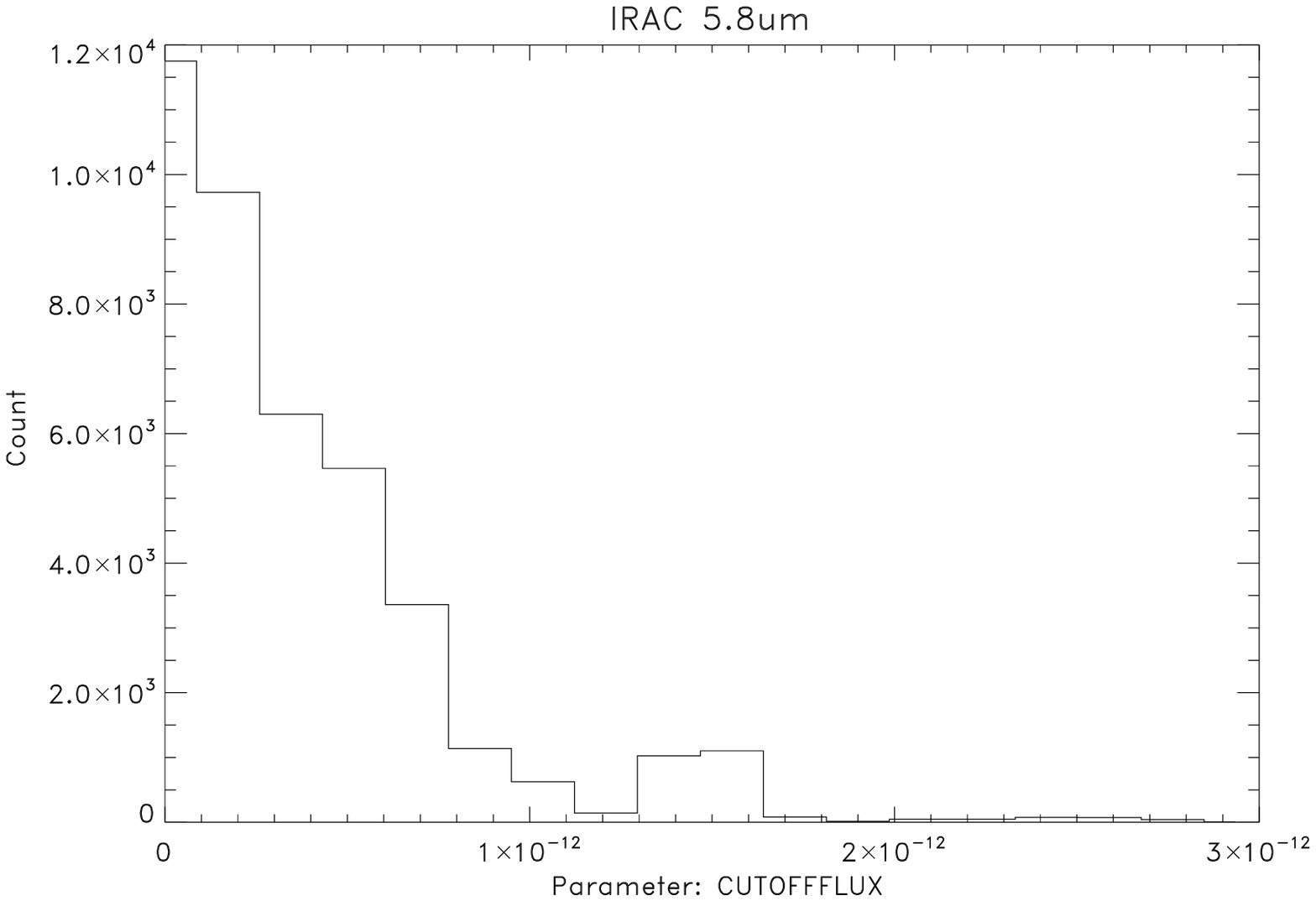 ,        angle=0, width=5cm}
\epsfig{file=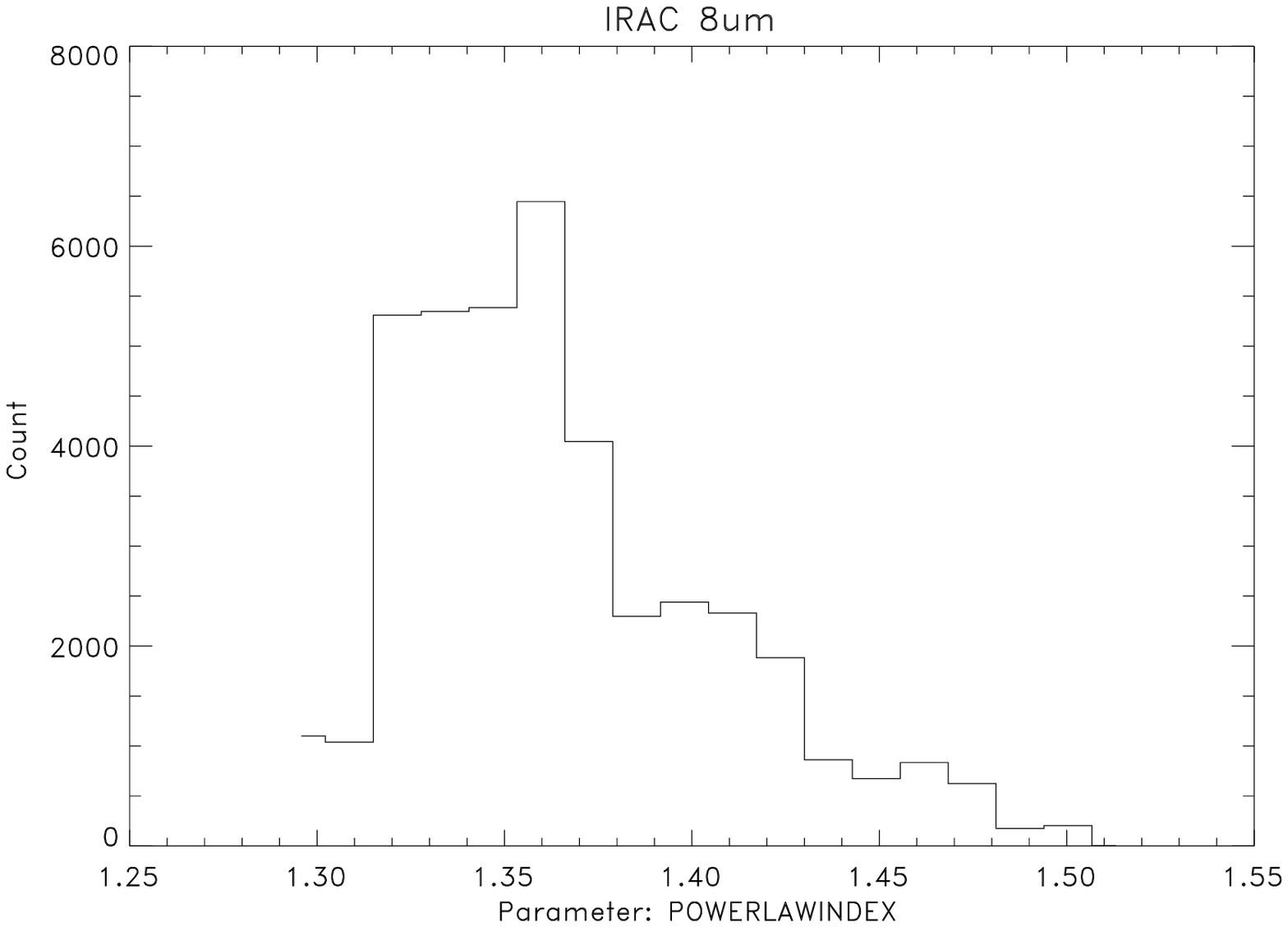 ,     angle=0, width=5cm}
\epsfig{file=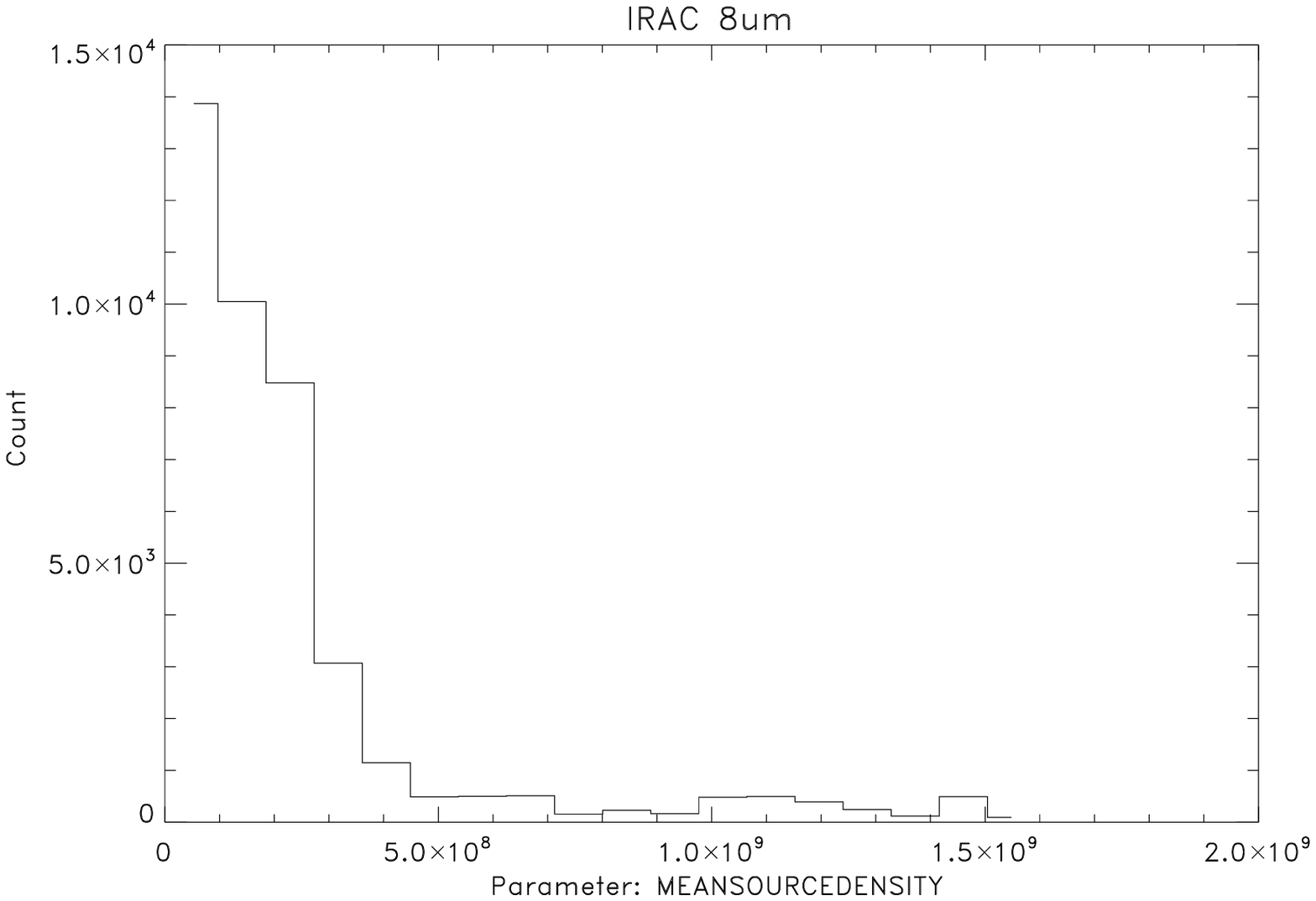 , angle=0, width=5cm}
\epsfig{file=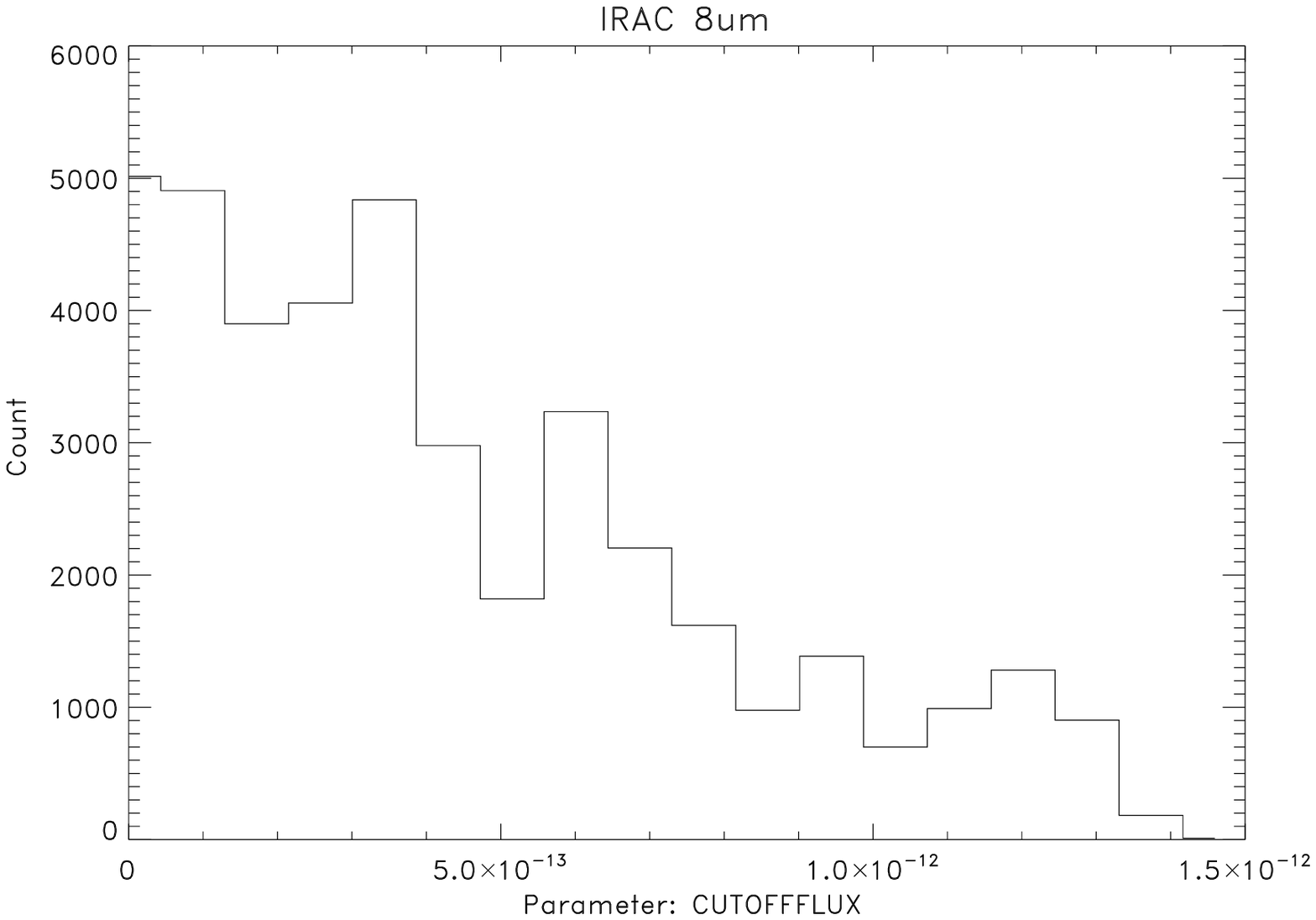 ,        angle=0, width=5cm}
\caption{The marginalised 1D posterior probability distributions for
  the fitted model parameters.  The distributions generated by taking
  a histogram of the MCMC samples and are given in units of number of
  counts per bin.  The level of structure in the distributions
  suggests that the MCMC chains have only borderline convergence
  (longer chains were not possible, due to computational
  constraints).  This will not have a substantial impact on the broad
  results of this analysis.}
\label{fig:1dLikelihoods}
\end{minipage}
\end{figure*}

One of the great advantages of MCMC sampling is that it enables us to
recover much information about the posterior probability distribution.
Figure \ref{fig:1dLikelihoods} shows the 1D, marginalised likelihoods
for each of the three fitted parameters, for each of the four IRAC
bands.  For any set of samples drawn from the posterior distribution,
marginalisation is straightforwardly achieved by taking a histogram of
the samples over the parameter of interest.

The constrained faint number count models can be integrated to find
estimates of their contribution to the the infra-red background
light.  Combining this with estimates of the contribution from bright
sources taken from \citet[][]{Fazio-04}, we can thus obtain overall
estimates of the integrated background light in the four IRAC bands.
We note that for the purposes of these analyses, we cut off our number
count models at the flux of the faintest Fazio et al point.

\begin{figure*}
\begin{minipage}{150mm}
\epsfig{file=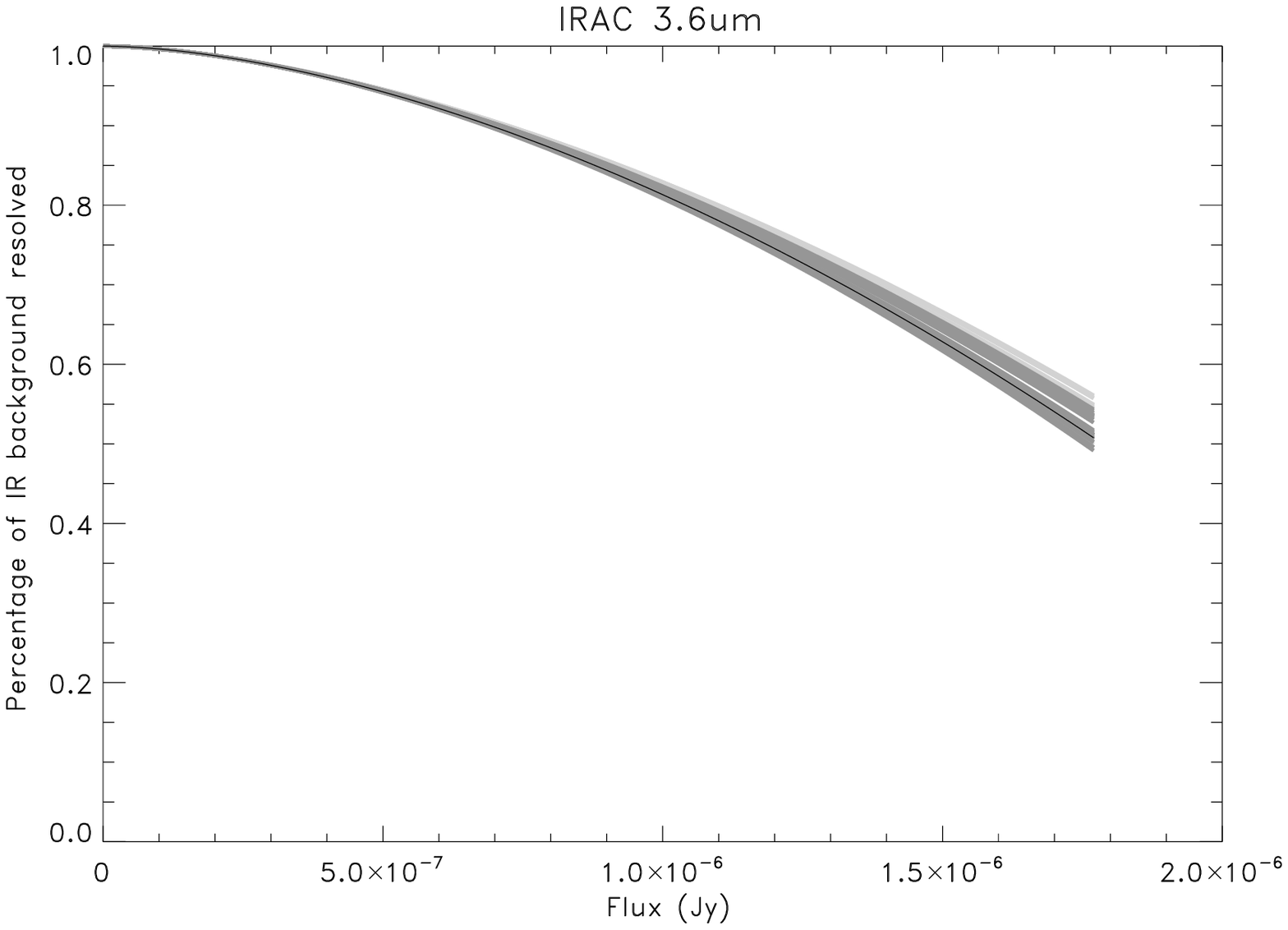  , angle=0, width=8.5cm}
\epsfig{file=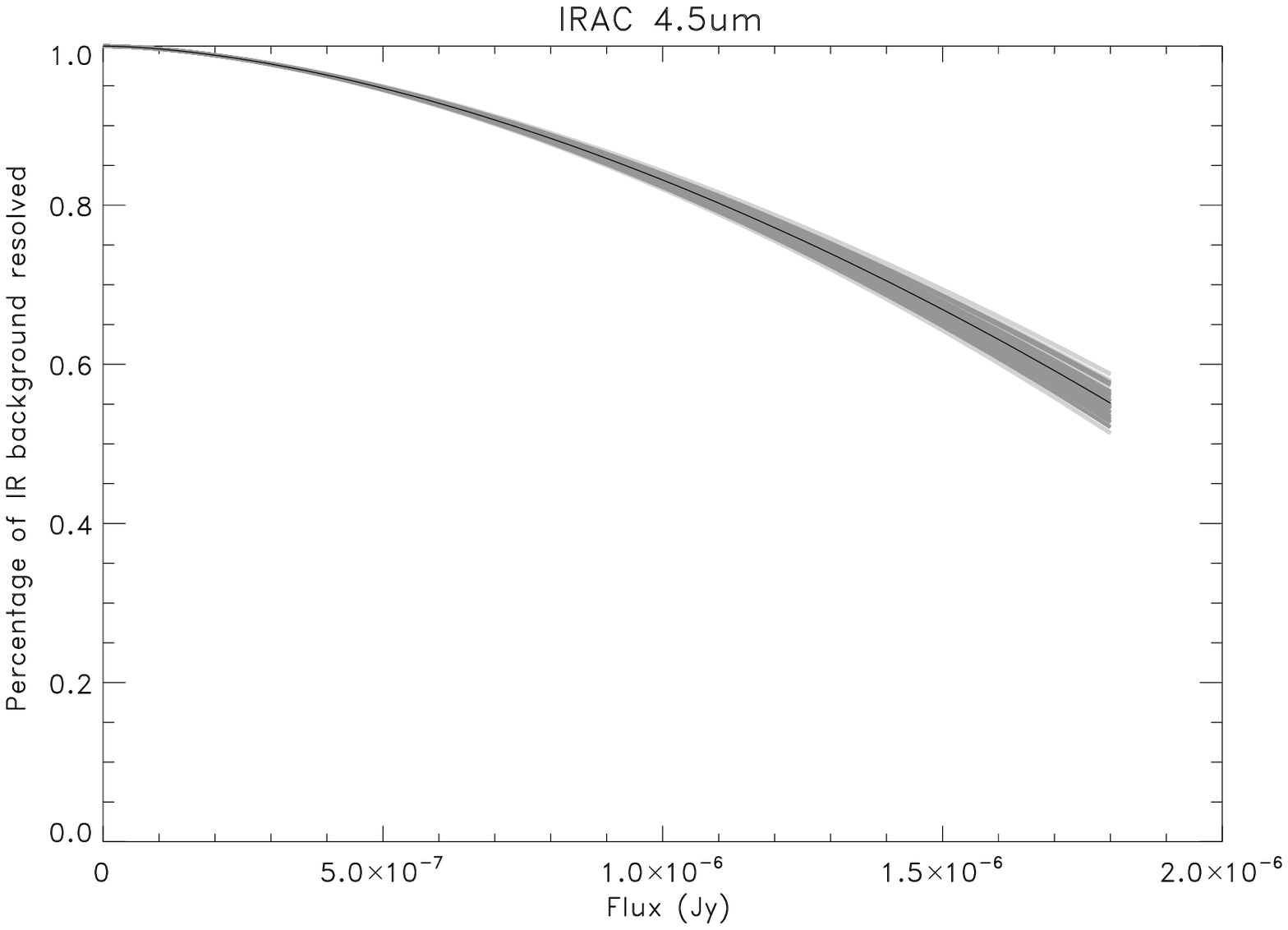  , angle=0, width=8.5cm}
\epsfig{file=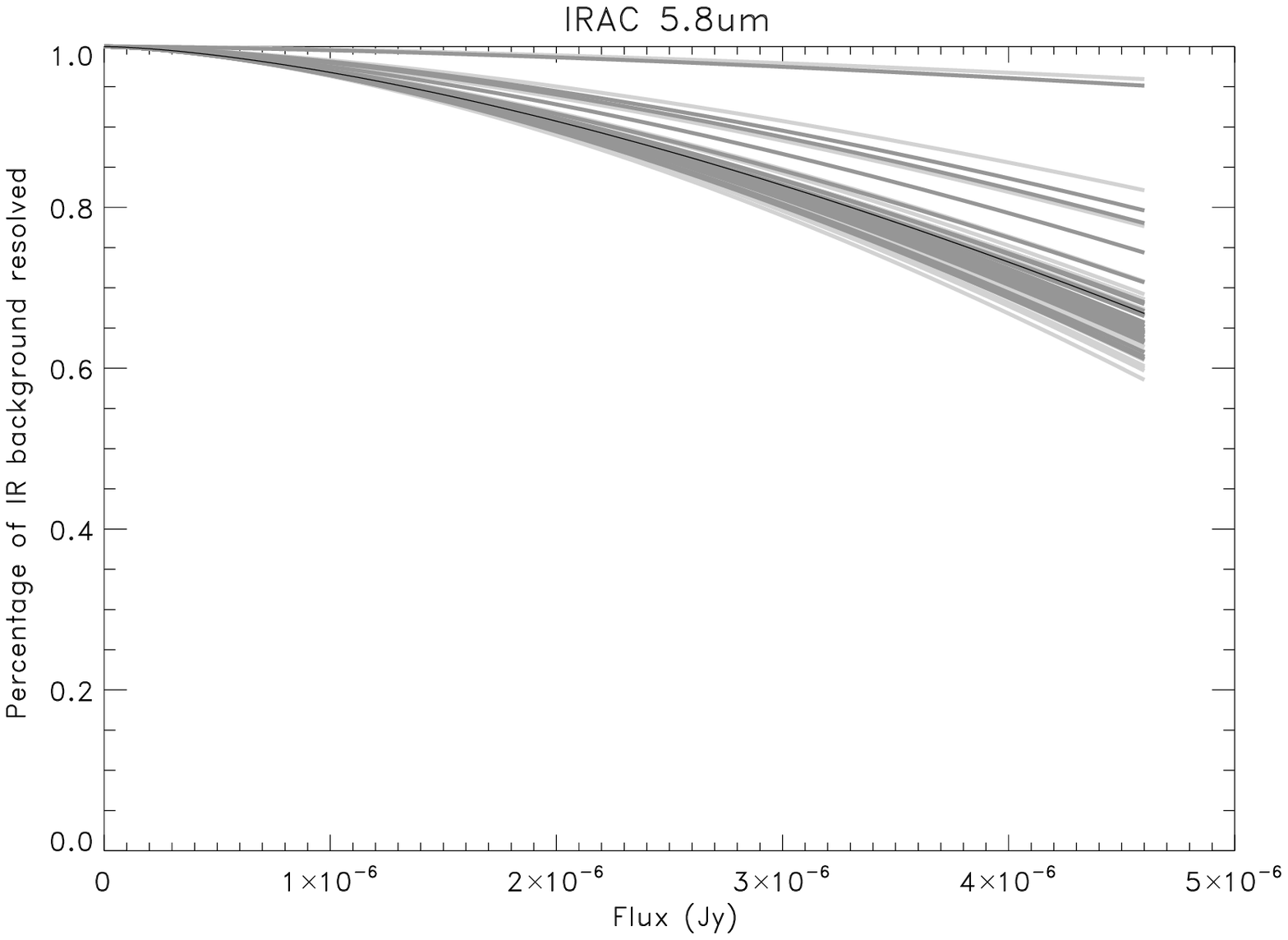  , angle=0, width=8.5cm}
\epsfig{file=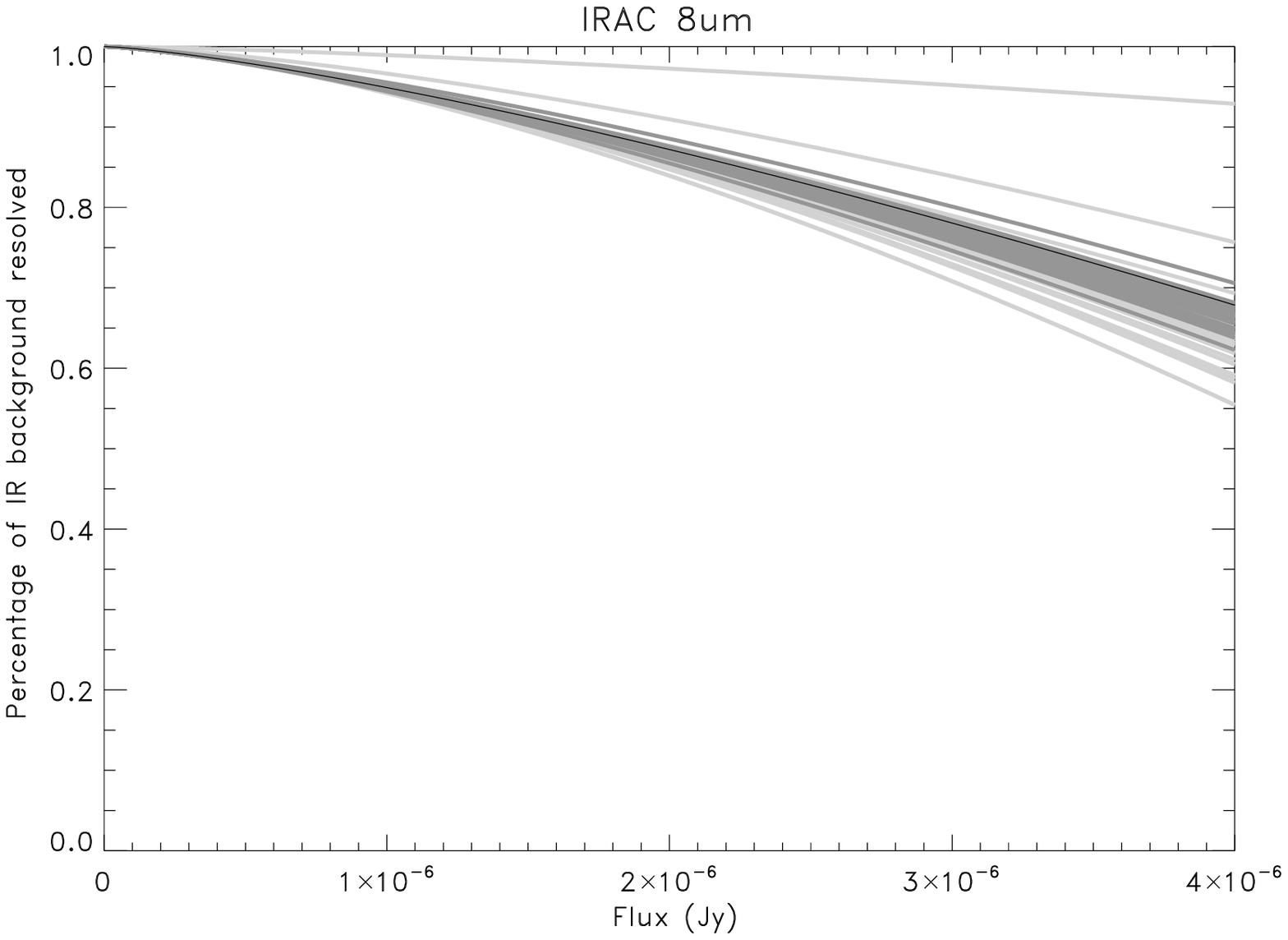  , angle=0, width=8.5cm}
\caption{Estimates of the proportion of the IR background light that
  is resolved, as a function of flux.  The proportion is calculated
  assuming that the 'total' IR background light is given by values
  from Table \ref{table:integratedBackground}.  Shown are the best-fit solution
(block line), along with a subset of the 68\% (dark grey) and 95\%
  (light grey) confidence region solutions.}
\label{fig:backgroundFraction}
\end{minipage}
\end{figure*}

Figure \ref{fig:backgroundFraction} shows the proportion of the total
background light resolved, as a function of flux.  The result for the
best-fit model is shown (black line); also plotted are a subset of the
68\% (dark grey) and 95\% (light grey) confidence region models, to
indicate the statistical uncertainty in the results.

Figure \ref{fig:1dLikelihoods} shows the marginalised 1D posterior
probability distributions for the total integrated background light
due to our constrained models (i.e. not including the contribution
from brighter galaxies).  These results are tabulated in table
\ref{table:integratedBackground} and combined with the contribution
from bright sources in order to provide estimates of the total
background light due to galaxies in the four IRAC bands.

\begin{figure*}
\begin{minipage}{150mm}
\epsfig{file=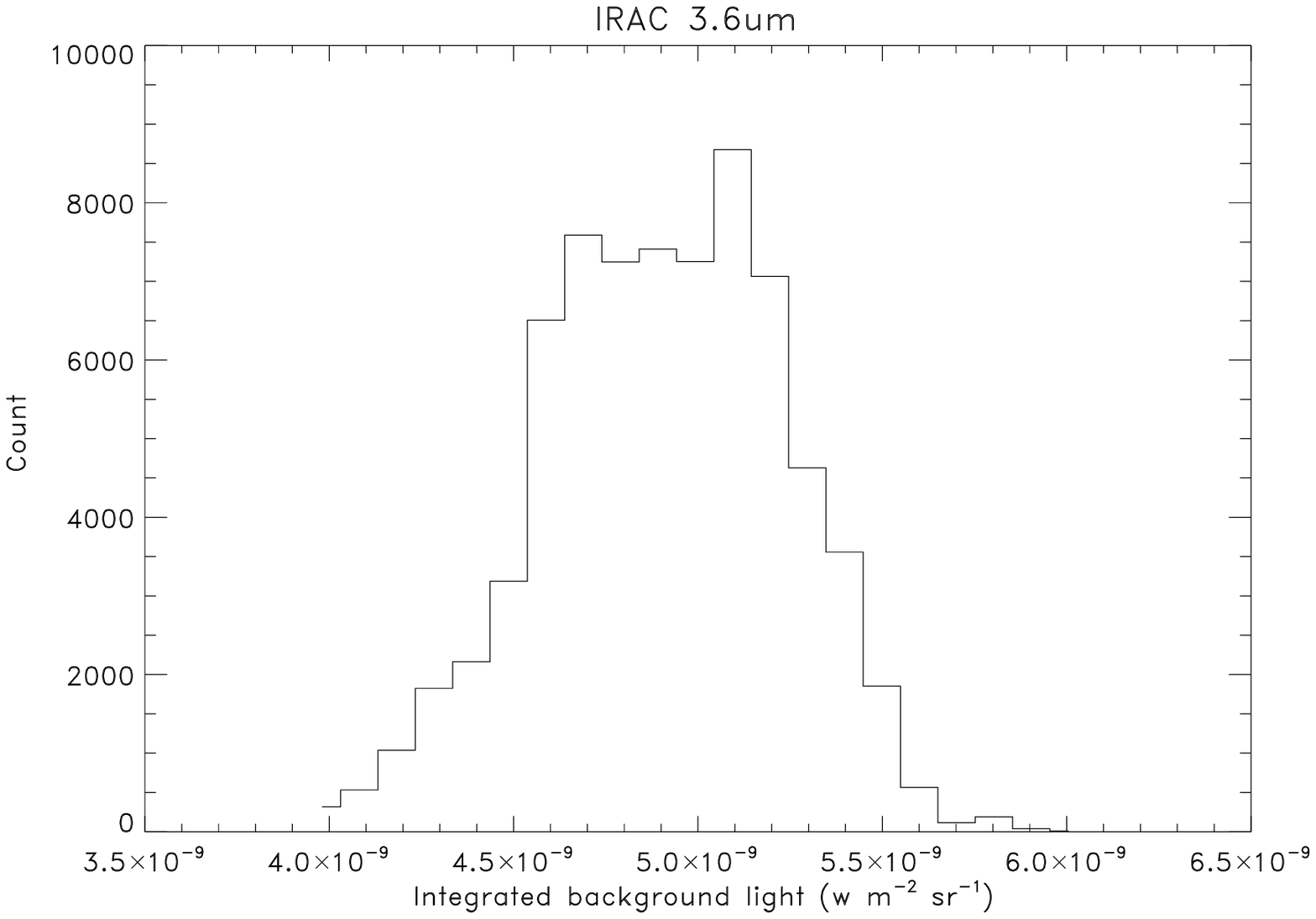  , angle=0, width=8.5cm}
\epsfig{file=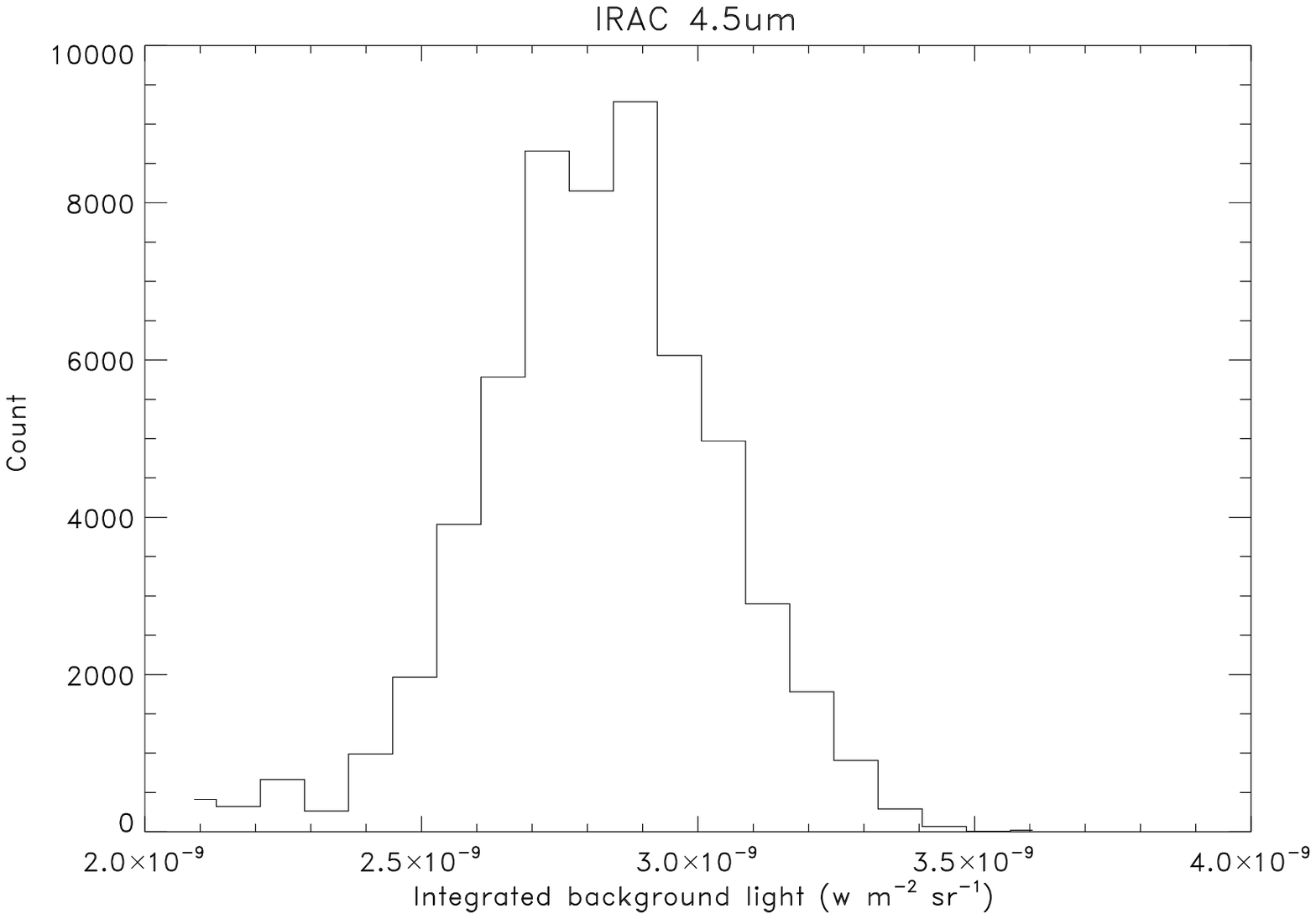  , angle=0, width=8.5cm}
\epsfig{file=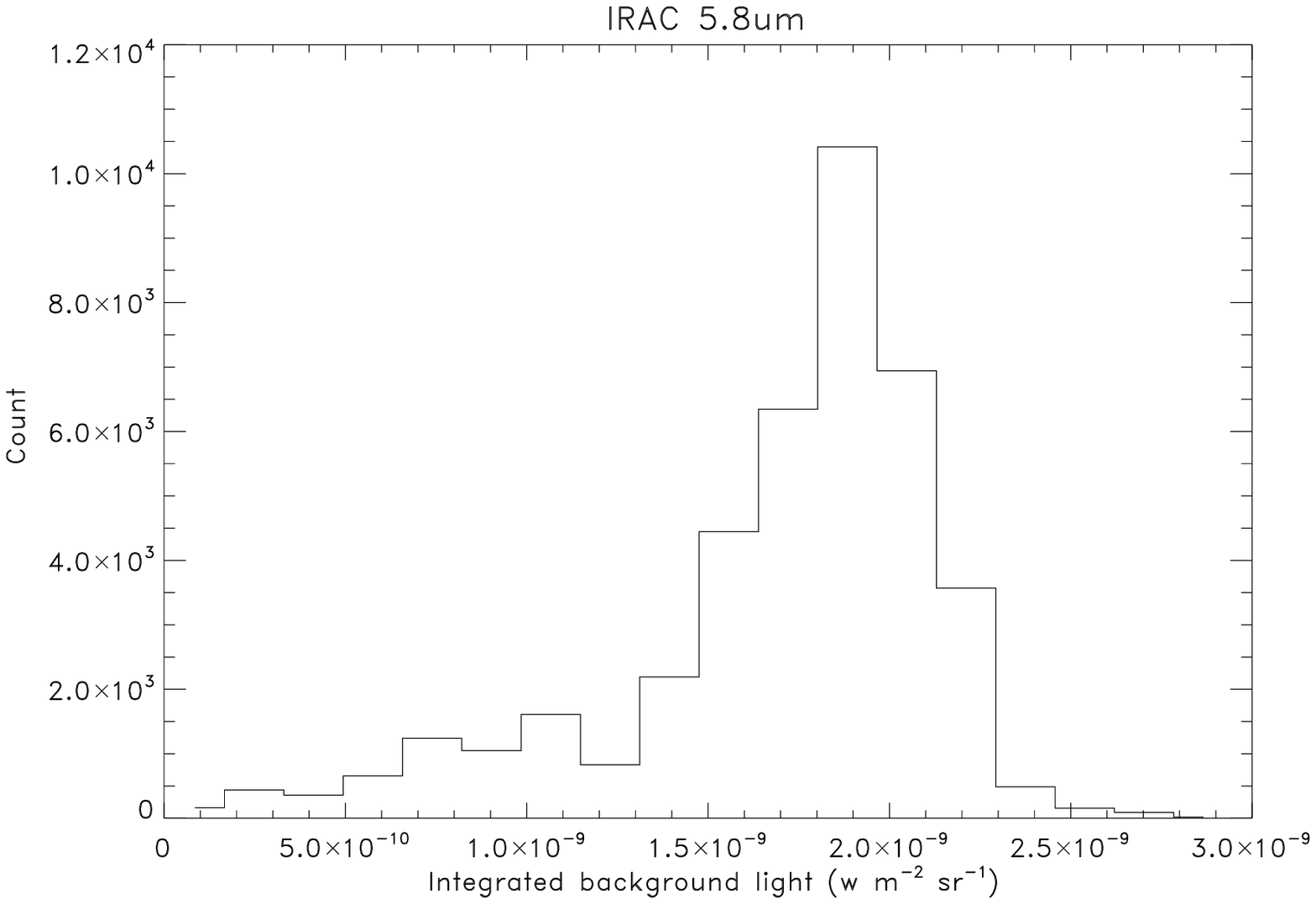  , angle=0, width=8.5cm}
\epsfig{file=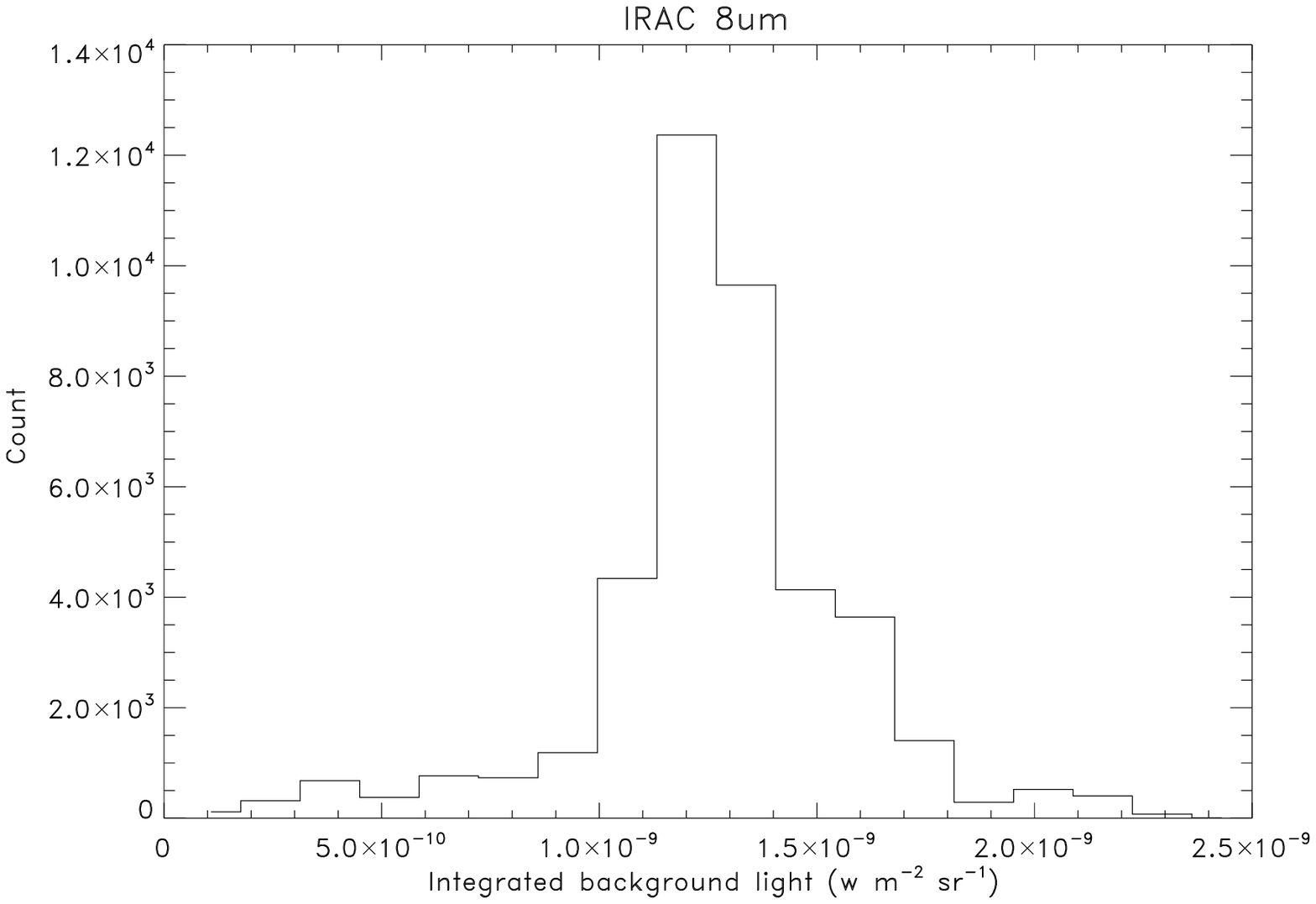  , angle=0, width=8.5cm}
\caption{The PDFs of the integrated background light due to the
  constrained number count models resulting from our analyses.  These
  estimates \emph{do not} include the contribution from brighter
  sources.  The overall estimated contribution from all sources is
  given in Table \ref{table:integratedBackground}.}
\label{fig:backgroundHistogram}
\end{minipage}
\end{figure*}
\section{Discussion and conclusions} \label{section:conclusions}

Our analyses of the Spitzer GOODS HDF-N epoch one data have produced
constraints on the faint source counts in the four IRAC bands (3.6,
4.5, 5.8, 8 microns).  \citet[][]{Fazio-04} estimate the contribution
of stars to the counts at such faint fluxes to be be negligible.  We
therefore can regard the results in Figure \ref{fig:numberCountFits}
as approximating the faint galaxy counts in these bands.  The analyses
presented in this paper provide meaningful constraints on these
differential number counts down to $10^{-8} Jy$, over two orders of
magnitude fainter in flux than the faintest published number counts.

The plotted number count points in Figure \ref{fig:numberCountFits}
are the result of source extraction from
other Spitzer observations, making them measurements that are independent of
the data used in this paper.  The most noteworthy
discrepancy is that the fluctuation analyses seem to produce number
counts that are systematically higher than those measured by
\citet[][]{Fazio-04}.  There are a number of possible explanations for
this effect.

The first is suggested by Figure \ref{fig:numberCountFluxScaled}.  A
relative difference between the flux calibration of the data used in
this paper and that of \citet[][]{Fazio-04} could make the results
consistent with one another.

Another possibility is that the number count model we have fitted is an
over-simplification.  Being strictly phenomenological, 
it would not be hugely surprising to find at least some deviations of
this nature.  Some non-power-law behaviour in
the real faint number counts (for example, a plateau) could cause such
an effect.  Clustering of the real sources could also have an impact,
as it is known that source clustering changes the expected fluctuation
signal.

There is also the possibility of fluctuations from non-point-like
sources, such as diffuse galactic emission.  Fluctuation analysis is
likely to be sensitive to such effects in the data, should they be
present in substantial amounts.

Balanced against this is the goodness-of-fit of the models (see the
reduced chi-squared values in Table \ref{table:onePowerLaw}).  These
show that in all cases except the 3.6 micron band, the fitted models
are good representations of the data.  Any statistically significant
deviation of the real number counts from a power law would show up as
an increase in the reduced chi-squared values.

The 3.6 micron data is fitted substantially more poorly by the model,
so it is possible that we have detected non-power-law behaviour in
this case.  Indeed, we would expect the 3.6 micron data to be most
sensitive to such deviations, as it has the lowest relative level of
instrumental noise (see Figure \ref{fig:dataPDFs})
However, the fact that this over-estimate (relative to the
Fazio counts) occurs in all four bands suggests that there is also
something else occurring.

Other systematic effects in the Spitzer GOODS data may also play a part.
Any non-Gaussianity in the noise could be significant, although the
outlier rejection used to reduce the GOODS data guards against this.
Issues such as the residuals present from flat-fielding may also have
an impact, although if they add Gaussian noise, we have shown our
analysis to be robust to this. We
have also assumed a constant point spread function across the whole
image, which may not be the case.  Effects of this nature may
contribute to an explanation of the observed discrepancy.

Similarly, there may be systematic effects present in the Fazio number
counts.  Faint source counts are, by their nature, very difficult to
measure to high precision, especially when the issue of incompleteness
is a factor (as it is here).

Although the data used in this paper has been background-subtracted,
the possibility of some residual contamination from diffuse
backgrounds such as galactic cirrus remains.  The exact
effect of this is unclear and it may be negligible.  However, it may
have some impact on our results.  

Overall, it is clear that we expect some level of systematic
uncertainty in addition to the more easily quantified random
uncertainty.  The level of consistency between the results from
fluctuation analyses and direct source counting give us a crude
measure of the level of these systematic uncertainties.

Figure \ref{fig:1dLikelihoods} shows the 1D marginalised likelihoods
for the fitted parameters.  Each of these histograms is formed from at
least 40,000 MCMC samples.  This number is a little low for MCMC
sampling analyses and this is reflected in some of the plots, where a
degree of multiple peak structure can be seen.  These features are more
likely to represent imperfect convergence of the MCMC chains than any
real feature, the result of the computational constraints on these
analyses.  Despite this, the broad results such as best-fit and
confidence regions are still valid.  

Figure \ref{fig:numberCountFits} shows our results in comparison to
the galaxy number count models of \citet[][]{Pearson-05} and
\citet[][]{Franceschini-05}. Of particular note is the large excess of faint number
counts in our results over those predicted the models.  This result
remains if the inconsistency between our results and the Fazio number
counts is explained by a difference in flux calibration.  If this is
the case, this result indicates the presence of a significant amount
of faint fluctuations in the IR background.  

Constraint of the faint number counts allows us to estimate the
contribution of faint galaxies to the integrated background
light.  One of the great advantages of MCMC analysis is that we can do
this for each MCMC sample, thus correctly propagating the statistical
uncertainty of our analyses to new quantities such as this.  The plots
in Figure \ref{fig:backgroundFraction} show that the Fazio counts
reach a depth sufficient to resolve of order half of more of the
integrated background light.  The fluctuation analysis presented in
this paper extends to fluxes over a factor of 100 fainter than this,
therefore covering virtually all of the integrated background light in
the four IRAC bands.

We have estimated the total integrated background light in each of the
four IRAC bands (see Figure \ref{fig:1dLikelihoods}  and Table
\ref{table:integratedBackground}).  The IRAC 3.6 micron band is comparable
to the near-IR L-band (2.9 to 3.5 microns), for which there are
several existing estimates of the integrated background light due to
galaxies.  In comparison to our 3.6 micron result of
$10.6^{+0.63}_{-1.95} \times 10^{-9} w m^{-2} sr^{-1}$,
\citet[][]{DwekArendt-98} obtain 
$9.9 \pm 2.9 \times 10^{-9} w m^{-2} sr^{-1}$, 
\citet[][]{Gorjian-00} obtain
$11.0 \pm 3.3 \times 10^{-9} w m^{-2} sr^{-1}$, 
and \citet[][]{WrightReese-00} find 
$12.4 \pm 3.2 \times 10^{-9} w m^{-2} sr^{-1}$.
We therefore conclude that our 3.6 micron results are consistent with
all three of these L-band results.  We also note that our errors are
substantially smaller than for these measurements.

Also of interest are the results of \citet[]{Matsumoto-05}, who
measure the near-IR integrated background light using the Infra-Red
Telescope in Space (IRTS).  Their data extend from 1 to 4 microns.  A
representative measurement they obtain of the integrated background is
(3.68 microns) 
$13.1 \pm 3.7 \times 10^{-9} w m^{-2} sr^{-1}$.  On the basis of our
3.6 micron results, we therefore suggest that the longer wavelength
results from \citet[]{Matsumoto-05} can be explained by the emission
of faint, point-like objects, as seen in the Spitzer GOODS data.  We
note that this is consistent with the interpretations
\citet[]{Matsumoto-05} have made using additional information from
power spectra and colour information.

We also note the recent results of \citet[][]{Kashlinsky-05}, who detect an
excess in the IRAC-band infra-red background, which they attribute to
population III stars.  While our analysis does not provide information
to specifically comment on this interpretation, it is instructive to
compare our findings.  \citet[][]{Kashlinsky-05} detect an excess in the
infra-red background after subtracting a contribution from faint
galaxies which they determine by extrapolating the
\citet[][]{Fazio-04} number counts using a fitted power law number
count model.  Our results likewise find a substantial excess over the
extrapolated Fazio number counts.  Furthermore, our results show a
large excess over the faint number count model predictions shown in
Figure \ref{fig:numberCountFits}.  Even if this were in part due to
calibration uncertainty, the differing power law slopes between our
results and the models mean that some excess will remain (see, for
example, Figure \ref{fig:numberCountFluxScaled}).

The analyses in this paper are subject to a number of caveats (which
we have discussed above).  We reach a number of conclusions (see below); these 
are all subject to the assumptions we have made in this analysis.  Model-wise, We
have assumed Gaussian instrumental noise (of known variance), a power
law number count model over the range of interest, and that the
sources are Poisson-distributed and point-like.  Data-wise, we use
the best available flux calibration and point spread function.  We
also assume that flat-fielding artifacts do not significantly bias
our results.  In conclusion, in this paper we present the following results.

\begin{enumerate}
\item{\tt Constraints on near-IR galaxy number counts.}  
  We use the Spitzer GOODS survey (HDF-N, data release one) to place
  constraints on the faint number counts in the four near-IR
  Spitzer IRAC wavebands.  As the contribution due to stars at these
  fluxes is predicted to be small, these number counts approximate to
  galaxy number counts.  We generate meaningful constraints on these
  number counts down to $10^{-8}Jy$, over two orders of magnitude
  fainter in flux than the deepest currently published number counts.
\item{\tt A discrepancy between our results and those of
  \citet[][]{Fazio-04}.}
  Our results systematically overestimate the differential number
  counts with respect to the Fazio counts.  We discuss a number of
  possible causes of this.
\item{\tt Determination of the background light due to unresolved
  point sources in each of the four IRAC bands.}
  We estimate the proportion of background light dues to faint
  galaxies that is resolved, as a function of flux.  We concur with
  the broad estimates of \citet[][]{Fazio-04} that over half the
  background light is resolved by Spitzer into individual galaxies.
  We demonstrate that almost the entirety of the remaining integrated
  light is probed by our fluctuation analysis.
  We also estimate the contribution of faint galaxies to the integrated
  background light.  By adding this to the estimate for bright sources
  given by \citet[][]{Fazio-04}, we obtain estimates of the total
  integrated background light.  Our 3.6 micron result is found to be
  consistent with the L-band results of \citet[][]{DwekArendt-98},
  \citet[][]{Gorjian-00} and \citet[][]{WrightReese-00}.  This result
  is also consistent with the results from \citet[]{Matsumoto-05} at
  the corresponding wavelength.  We note that our analyses do not
  impact on the substantial excess detected by \citet[]{Matsumoto-05}
  at shorter wavelengths.
\item{\tt Consistency with the recent detection of excess infra-red
  background reported by \citet[][]{Kashlinsky-05}.}
  Although our analyses do not provide information of the
  interpretation of the \citet[][]{Kashlinsky-05} result, we do detect a
  clear excess over the \citet[][]{Fazio-04} number counts; the
  \citet[][]{Kashlinsky-05} is also based on the existence of such an excess.
\end{enumerate}

Fluctuation analyses are an invaluable way of probing galaxy
number count distributions well below the confusion limit of any given
instrument.  Photometric surveys are now producing data sets where the
statistical errors on such analyses are sub-dominant to the systematic
uncertainties on such measurements.  To improve the knowledge we can
extract from fluctuation analyses, we must therefore pay particular
attention to both understanding and accounting for the systematic
influences on our data.

\section*{Acknowledgments}
Richard Savage thanks the Particle Physics and Astronomy Research
Council (PPARC) for support under grant PPA/G/S/2002/00481.

Seb Oliver thanks the Leverhulme Trust for support in the form of a
Leverhulme research fellowship.

We thank Chris Pearson, Alberto Franceschini and Giulia Rodighiero for
their kind provision of the number count models shown in Figure
\ref{fig:numberCountFits}.

This work is based on observations made with the Spitzer
Space Telescope, which is operated by the Jet Propulsion Laboratory,
California Institute of Technology under a contract with NASA.  

We particularly acknowledge the IRAC instrument and the Spitzer GOODS
legacy survey.

\label{lastpage}
\bibliography{../bibtex_files/pd_analysis_refs,../bibtex_files/inference_refs,../bibtex_files/Spitzer_refs,../bibtex_files/IR_refs,../bibtex_files/ASTRO-F_refs,../bibtex_files/CIRB_refs}

\begin{thebibliography}{}

\bibitem[\protect\citeauthoryear{{Barcons}, {Raymont}, {Warwick}, {Fabian},
  {Mason}, {McHardy} \& {Rowan-Robinson}}{{Barcons}
  et~al.}{1994}]{XrayPDAnalysis-94}
{Barcons} X.,  {Raymont} G.~B.,  {Warwick} R.~S.,  {Fabian} A.~C.,  {Mason}
  K.~O.,  {McHardy} I.,    {Rowan-Robinson} M.,  1994, \mnras, 268, 833

\bibitem[\protect\citeauthoryear{{Condon}}{{Condon}}{1974}]{Condon-74}
{Condon} J.~J.,  1974, \apj, 188, 279

\bibitem[\protect\citeauthoryear{{Craig}, {Mendez} \& {Wright}}{{Craig}
  et~al.}{2004}]{WISE-2004}
{Craig} N.,  {Mendez} B.~J.,    {Wright} E.~L.,  2004, American Astronomical
  Society Meeting Abstracts, 205,

\bibitem[\protect\citeauthoryear{{Dickinson} \& {GOODS}}{{Dickinson} \&
  {GOODS}}{2004}]{SpitzerGOODS-2004}
{Dickinson} M.,  {GOODS} 2004, American Astronomical Society Meeting Abstracts,
  205,

\bibitem[\protect\citeauthoryear{{Dwek} \& {Arendt}}{{Dwek} \&
  {Arendt}}{1998}]{DwekArendt-98}
{Dwek} E.,  {Arendt} R.~G.,  1998, \apjl, 508, L9

\bibitem[\protect\citeauthoryear{{Fazio}, {Ashby}, {Barmby}, {Hora}, {Huang},
  {Pahre}, {Wang}, {Willner}, {Arendt}, {Moseley}, {Brodwin}, {Eisenhardt},
  {Stern}, {Tollestrup} \& {Wright}}{{Fazio} et~al.}{2004}]{Fazio-04}
{Fazio} G.~G.,  {Ashby} M.~L.~N.,  {Barmby} P.,  {Hora} J.~L.,  {Huang} J.-S.,
  {Pahre} M.~A.,  {Wang} Z.,  {Willner} S.~P.,  {Arendt} R.~G.,  {Moseley}
  S.~H.,  {Brodwin} M.,  {Eisenhardt} P.,  {Stern} D.,  {Tollestrup} E.~V.,
  {Wright} E.~L.,  2004, \apjs, 154, 39

\bibitem[\protect\citeauthoryear{{Fazio}}{{Fazio}}{2004}]{SpitzerIRAC-2004}
{Fazio} G.~G. e.~a.,  2004, \apjs, 154, 10

\bibitem[\protect\citeauthoryear{{Franceschini} et~al.,}{{Franceschini}
  et~al.}{2005}]{Franceschini-05}
{Franceschini} A.,  et~al., 2005, submitted

\bibitem[\protect\citeauthoryear{{Franceschini}, {Martin-Mirones}, {Danese} \&
  {de Zotti}}{{Franceschini} et~al.}{1993}]{XrayPDAnalysis-93}
{Franceschini} A.,  {Martin-Mirones} J.~M.,  {Danese} L.,    {de Zotti} G.,
  1993, \mnras, 264, 35

\bibitem[\protect\citeauthoryear{{Gilks}, {Richardson} \&
  {Spiegelhalter}}{{Gilks} et~al.}{1995}]{GilksMCMC-book}
{Gilks} W.~R.,  {Richardson} S.,    {Spiegelhalter} D.~J.,  1995, Marko Chain
  Monte Carlo in practice.
Chapman \& Hall, London

\bibitem[\protect\citeauthoryear{{Gorjian}, {Wright} \& {Chary}}{{Gorjian}
  et~al.}{2000}]{Gorjian-00}
{Gorjian} V.,  {Wright} E.~L.,    {Chary} R.~R.,  2000, \apj, 536, 550

\bibitem[\protect\citeauthoryear{{Kashlinsky}, {Arendt}, {Mather} \&
  {Moseley}}{{Kashlinsky} et~al.}{2005}]{Kashlinsky-05}
{Kashlinsky} A.,  {Arendt} R.~G.,  {Mather} J.,    {Moseley} S.~H.,  2005,
  \nat, 438

\bibitem[\protect\citeauthoryear{{Matsumoto}, {Matsuura}, {Murakami}, {Tanaka},
  {Freund}, {Lim}, {Cohen}, {Kawada} \& {Noda}}{{Matsumoto}
  et~al.}{2005}]{Matsumoto-05}
{Matsumoto} T.,  {Matsuura} S.,  {Murakami} H.,  {Tanaka} M.,  {Freund} M.,
  {Lim} M.,  {Cohen} M.,  {Kawada} M.,    {Noda} M.,  2005, \apj, 626, 31

\bibitem[\protect\citeauthoryear{{Nakagawa}}{{Nakagawa}}{2004}]{SPICA-2004}
{Nakagawa} T.,  2004, Advances in Space Research, 34, 645

\bibitem[\protect\citeauthoryear{{Oliver}}{{Oliver}}{1997}]{ISOpdAnalysis-97}
{Oliver} S.~J. e.~a.,  1997, \mnras, 289, 471

\bibitem[\protect\citeauthoryear{{Pearson}}{{Pearson}}{2005}]{Pearson-05}
{Pearson} C.,  2005, \mnras, 358, 1417

\bibitem[\protect\citeauthoryear{{Pearson}, {Shibai}, {Matsumoto}, {Murakami},
  {Nakagawa}, {Kawada}, {Onaka}, {Matsuhara}, {Kii}, {Yamamura} \&
  {Takagi}}{{Pearson} et~al.}{2004}]{ASTROFsuperIRAS-2004}
{Pearson} C.~P.,  {Shibai} H.,  {Matsumoto} T.,  {Murakami} H.,  {Nakagawa} T.,
   {Kawada} M.,  {Onaka} T.,  {Matsuhara} H.,  {Kii} T.,  {Yamamura} I.,
  {Takagi} T.,  2004, \mnras, 347, 1113

\bibitem[\protect\citeauthoryear{{Pilbratt}}{{Pilbratt}}{2004}]{HerschelOvervi%
ew-2004}
{Pilbratt} G.~L.,  2004, American Astronomical Society Meeting Abstracts, 204,

\bibitem[\protect\citeauthoryear{{Scheuer}}{{Scheuer}}{1957}]{Scheuer-57}
{Scheuer} P.,  1957, Proc. Cambridge Phil. Soc., 53, 764

\bibitem[\protect\citeauthoryear{{Scheuer}}{{Scheuer}}{1974}]{Scheuer-74}
{Scheuer} P.~A.~G.,  1974, \mnras, 166, 329

\bibitem[\protect\citeauthoryear{{Scott}}{{Scott}}{1992}]{Scott_densityEstimat%
ion}
{Scott} D.~W.,  1992, {Multivariate Density Estimation}.
Multivariate Density Estimation, Wiley, New York, 1992

\bibitem[\protect\citeauthoryear{{Vio}, {Fasano}, {Lazzarin} \& {Lessi}}{{Vio}
  et~al.}{1994}]{Vio-94}
{Vio} R.,  {Fasano} G.,  {Lazzarin} M.,    {Lessi} O.,  1994, \aap, 289, 640

\bibitem[\protect\citeauthoryear{{Wall}, {Scheuer}, {Pauliny-Toth} \&
  {Witzel}}{{Wall} et~al.}{1982}]{radioPDAnalysis-82}
{Wall} J.~V.,  {Scheuer} P.~A.~G.,  {Pauliny-Toth} I.~I.~K.,    {Witzel} A.,
  1982, \mnras, 198, 221

\bibitem[\protect\citeauthoryear{{Wright} \& {Reese}}{{Wright} \&
  {Reese}}{2000}]{WrightReese-00}
{Wright} E.~L.,  {Reese} E.~D.,  2000, \apj, 545, 43

\end{thebibliography}
\bibliographystyle{mn2e}
\bsp
\end{document}